\documentclass[
journal=jacsat,
manuscript=article]{achemso}

\usepackage{blindtext}
\usepackage{graphicx}
\usepackage{dcolumn}
\usepackage{bm}
\usepackage{siunitx}
\usepackage{color}
\usepackage{xcolor}
\usepackage{gensymb}
\usepackage{natbib}
\usepackage{mathtools}
\usepackage{amssymb}
\usepackage{setspace}

\author{Malte Kremser}
\affiliation{Walter Schottky Institut, Physik-Department and MCQST, Technische Universit\"at M\"unchen, Am Coulombwall 4, 85748 Garching, Germany}
\email{Malte.Kremser@wsi.tum.de}

\author{Mauro Brotons-Gisbert}
\affiliation{Institute for Photonics and Quantum Sciences, SUPA, Heriot-Watt University, Edinburgh EH14 4AS, UK}

\author{Johannes Kn\"orzer}
\affiliation{Max-Planck-Institute for Quantum Optics and MCQST, Hans-Kopfermann-Str. 1, 85748 Garching, Germany}

\author{Janine G\"uckelhorn}
\affiliation{Walter Schottky Institut, Physik-Department and MCQST, Technische Universit\"at M\"unchen, Am Coulombwall 4, 85748 Garching, Germany}

\author{Moritz Meyer}
\affiliation{Walter Schottky Institut, Physik-Department and MCQST, Technische Universit\"at M\"unchen, Am Coulombwall 4, 85748 Garching, Germany}

\author{Matteo Barbone}
\affiliation{Walter Schottky Institut, Department of Electrical and Computer Engineering and MCQST, Technische Universit\"at M\"unchen, Am Coulombwall 4, 85748 Garching, Germany}

\author{Andreas V. Stier}
\affiliation{Walter Schottky Institut, Physik-Department and MCQST, Technische Universit\"at M\"unchen, Am Coulombwall 4, 85748 Garching, Germany}

\author{Brian D. Gerardot}
\affiliation{Institute for Photonics and Quantum Sciences, SUPA, Heriot-Watt University, Edinburgh EH14 4AS, UK}

\author{Kai M\"uller}
\affiliation{Walter Schottky Institut, Department of Electrical and Computer Engineering and MCQST, Technische Universit\"at M\"unchen, Am Coulombwall 4, 85748 Garching, Germany}

\author{Jonathan J. Finley}
\affiliation{Walter Schottky Institut, Physik-Department and MCQST, Technische Universit\"at M\"unchen, Am Coulombwall 4, 85748 Garching, Germany}
\email{Jonathan.Finley@wsi.tum.de}

\title{Discrete Interactions between a few Interlayer Excitons Trapped at a MoSe$_2$-WSe$_2$~Heterointerface}

\begin{document}
\singlespacing

\maketitle


\begin{abstract}
\textit{Inter-}layer excitons (IXs) in hetero-bilayers of transition metal dichalcogenides (TMDs)\cite{Rivera2015,Wang2017,Nagler2017,Wilson2018,Zhang2017,Seyler2018,Tran2019,Schinner2013} represent an exciting emergent class of long-lived dipolar composite bosons in an atomically thin, near-ideal two-dimensional (2D) system\cite{Nagler2017,Laikhtman2009,Schinner2013}. The long-range interactions that arise from the spatial separation of electrons and holes can give rise to novel quantum, as well as classical multi-particle correlation effects. Indeed, first indications of exciton condensation \cite{Wang2019,Sigl2020} have been reported recently. In order to acquire a detailed understanding of the possible many-body effects, the fundamental interactions between individual IXs have to be studied. Here, we trap a tunable number of dipolar IXs ($N_{IX}\sim1-5$) within a nanoscale confinement potential induced by placing a MoSe$_2$-WSe$_2$ hetero-bilayer (HBL) onto an array of SiO$_2$ nanopillars\cite{Branny2017,Palacios-Berraquero2017}. We control the mean occupation of the IX trap via the optical excitation level and observe discrete sharp-line emission from different configurations of interacting IXs. The intensities of these features exhibit characteristic near linear, quadratic, cubic, quartic and quintic power dependencies,\cite{Finley2001} which allows us to identify them as different multiparticle configurations with $N_{IX}\sim1-5$. We directly measure the hierarchy of dipolar and exchange interactions as $N_{IX}$ increases. The interlayer \textit{biexciton} ($N_{IX}=2$) is found to be an emission doublet that is blue-shifted from the single exciton by $\Delta E=\SI{8.4\pm0.6}{\milli\electronvolt}$ and split by $2J=\SI{1.2\pm0.5}{\milli\electronvolt}$. The blueshift is even more pronounced for triexcitons (\SI{12.4\pm0.4}{\milli\electronvolt}), quadexcitons (\SI{15.5\pm0.6}{\milli\electronvolt}) and quintexcitons (\SI{18.2\pm0.8}{\milli\electronvolt}). These values are shown to be mutually consistent with numerical modelling of dipolar excitons confined to a harmonic trapping potential having a confinement lengthscale in the range $\ell\approx \SI{3}{\nano\meter}$.\cite{Kylanpaa2015,Bondarev2018} Our results contribute to the understanding of interactions between IXs in TMD hetero-bilayers at the discrete limit of only a few excitations and represent a key step towards exploring quantum correlations between IXs in TMD hetero-bilayers.
\end{abstract}

\section{Introduction}

HBLs of van-der-Waals bonded 2D TMDs feature an atomically sharp interface with a type-II band alignment and frontier orbital couplings defined by the mutual angular orientation of the basal plane of the component layers.\cite{Rivera2015} Such atomically thin 2D heterointerfaces can host IXs \textendash {} Coulomb-bound states of electrons ($e$) and holes ($h$) located in the different component monolayers.\cite{Rivera2015} The spatial separation of $e$ and $h$ gives rise to a significantly increased lifetime compared to \textit{intra}layer excitons \cite{Rivera2015,Nagler2017,Miller2017} and a static electric out-of-plane dipole moment with a magnitude $\vert p_z\vert/e\sim \SI{0.7}{\nano\meter}$ for the MoSe$_2$-WSe$_2$ heterointerface\cite{Rivera2015}. The enhanced radiative lifetime of IXs allows them to cool toward the lattice temperature before recombination takes place, whereas the large static dipole moment facilitates tuning of the IX energy\cite{Rivera2015,Kiemle2018,Jauregui2018} and lifetime \cite{Kiemle2018,Jauregui2018} using static electric fields. In optical experiments, dipole-dipole repulsion between IXs gives rise to strongly blueshifting emission with increasing IX density and the potential to explore the many-body physics of dipolar composite bosons in a solid-state system\cite{Rivera2015,Nagler2017,Miller2017}. It has also been demonstrated that applying strain to the HBL can strongly tune the IX energy\cite{He2016}; a method that is routinely used to define exciton trapping potentials in monolayer TMDs. \cite{Branny2017,Palacios-Berraquero2017,Branny2016} Overall, IXs constitute a highly tunable platform for optically exploring and controlling interacting gases of dipolar excitons in the solid state.


\section{Results and Discussion}

\subsection*{Trapping interlayer excitons}


\begin{figure*}
\includegraphics{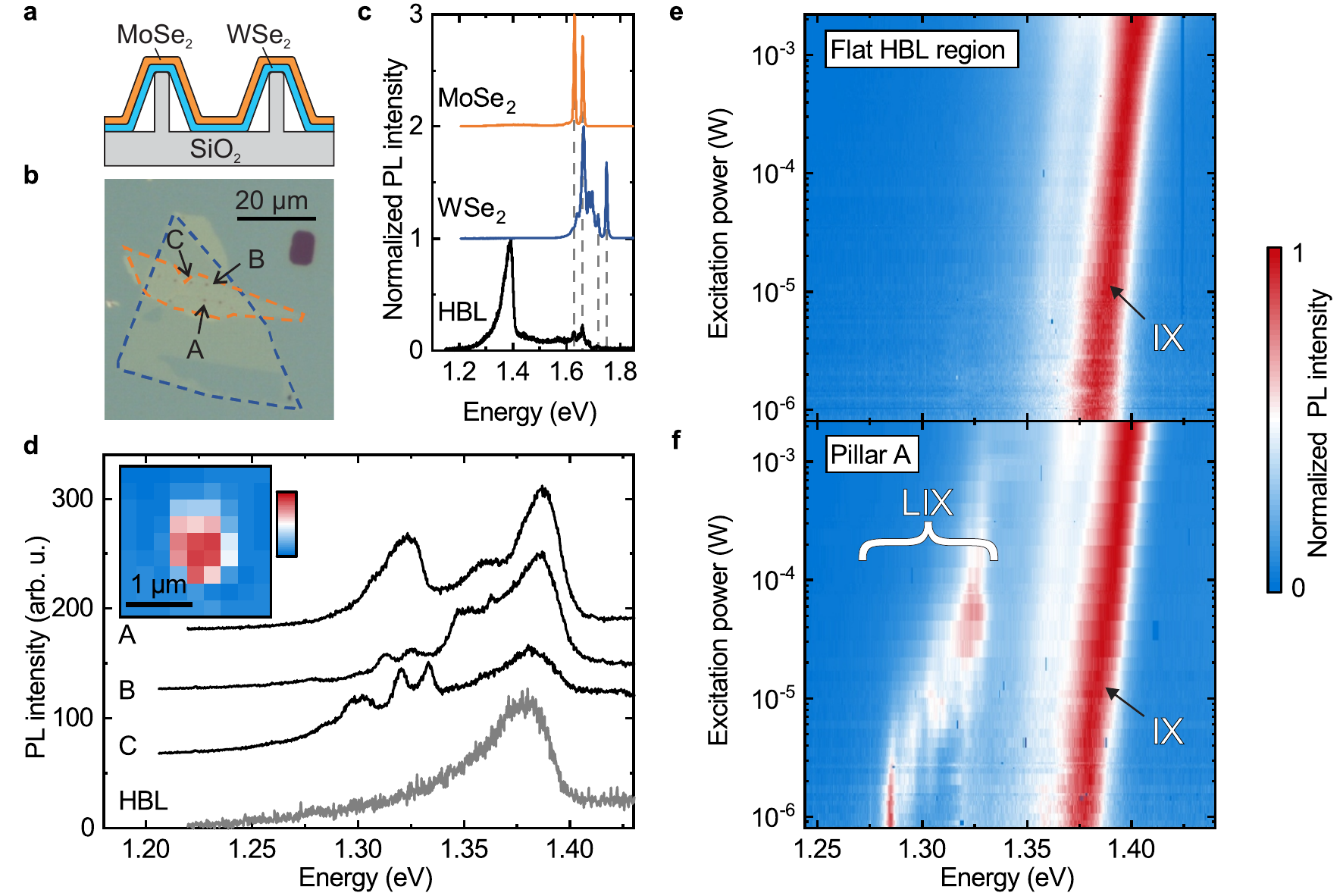}
\caption{\textbf{Low-temperature photoluminescence from free and localized interlayer excitons in a MoSe$_2$-WSe$_2$ heterobilayer.} \textbf{a} Schematic represenation, and \textbf{b} optical micrograph of the sample used in this work. The monolayers of WSe$_2$ and MoSe$_2$ are outlined with blue and orange dashed lines, respectively. The nanopillars appear as black dots in the heterobilayer region. \textbf{c} Example spectra recorded from the unstrained monolayer and heterobilayer regions. \textbf{d} Typical $\mu PL$ spectra recorded from three pillar sites (A, B and C as marked in a) in the heterobilayer region and a reference spectrum from the unstrained heterobilayer. Inset: Example spatially resolved integrated PL intensity of a low-energy emission peak at pillar~A. \textbf{e,f} Intensity-normalized PL spectra recorded from the HBL at different excitation powers at a position outside the pillar region (e) and on pillar~A (f). All spectra were recorded at \SI{10}{\kelvin} and, besides the power-dependent measurements shown in panels e and f, subject to strong continuous wave (cw) excitation at \SI{633}{\nano\meter} using a power of \SI{35}{\micro\watt} focused to a spot-size of $\sim \SI{1}{\micro\meter}$.} 
\label{fig:IX}
\end{figure*}

Our sample structure is schematically depicted in Fig.~\ref{fig:IX}a.  It consists of stacked monolayers of WSe$_2$ and MoSe$_2$ that form a HBL which is then transferred onto a SiO$_2$ substrate patterned with nanopillars having a height of $\sim \SI{90}{\nano\meter}$ and diameter $\sim \SI{130}{\nano\meter}$ (see Methods). Figure~\ref{fig:IX}b shows an optical micrograph of the sample in which the monolayer MoSe$_2$, WSe$_2$ and HBL regions can be seen, as well as the nanopillars in the HBL region. The contours of the WSe$_2$ and MoSe$_2$ monolayers are outlined with blue and orange dashed lines, respectively, and pillars that did not pierce the TMD layers appear as black dots. Figure~\ref{fig:IX}c shows normalized spectra recorded from three different unstrained sample regions, namely the WSe$_2$ monolayer, the MoSe$_2$ monolayer and the WSe$_2$-MoSe$_2$-HBL away from the nanopillars. These spectra were recorded at $T=\SI{10}{\kelvin}$ with cw excitation at \SI{633}{\nano\meter} and a power of $P_{ex}$=\SI{35}{\micro\watt} focused to $\sim \SI{1}{\micro\meter}$.
Both, WSe$_2$ and MoSe$_2$, have prominent emission features at their characteristic exciton and trion energies at \SI{1.75}{\electronvolt} and \SI{1.72}{\electronvolt}, \cite{Jones2013} and \SI{1.66}{\electronvolt} and \SI{1.63}{\electronvolt}, \cite{Ross2013} respectively. Furthermore, the WSe$_2$ component monolayer exhibits a broader band of unresolved emission at lower energies (\SIrange{1.60}{1.70}{\electronvolt}) that has been identified as stemming from dark excitons \cite{Wang2017a,Zhou2017} and trions, \cite{Zhang2017b} charged biexcitons \cite{Chen2018,Ye2018,Barbone2018,Li2018,Stevens2018} and defect-bound excitons.\cite{Tongay2013,Zhang2017a} Each of these emission features are also present in the HBL region, albeit strongly quenched, due to interlayer charge transfer that occurs over sub-picosecond timescales, faster than the exciton lifetime.\cite{Ceballos2014,Ceballos2015,Rigosi2016} The most prominent emission feature from the HBL is at $\sim$\SI{1.38}{\electronvolt} and is attributed to IXs formed by an electron in the MoSe$_2$ and a hole in the WSe$_2$. \cite{Rivera2015} This attribution is supported by the emission energy and the characteristically asymmetric lineshape featuring red-shifted emission from momentum-indirect IXs\cite{Miller2017}. Figure~\ref{fig:IX}e shows the power dependence of the IX emission revealing a significant blueshift ($\geq \SI{20}{\milli\electronvolt}$) arising from repulsive dipolar interactions in the gas of IX for increasing excitation levels\cite{Laikhtman2009,Nagler2017}. We note that this blueshift does \textit{not} occur over the whole range of excitation levels, but rather exhibits an onset close to $P_{ex}\sim \SI{30}{\nano\watt}$, marking a low-density regime where dipolar interactions are weak (see Suppl.~Fig.~2). Furthermore, the low-density IX regime also manifests itself by a deviation from the $\sqrt{P_{ex}}$ dependence of the IX emission intensity, which might be indicative of non-radiative exciton-exciton annihilation\cite{Amani2015} only above a threshold density.

The nanopillars have a strong impact on the IX emission spectra.  Typical data is presented in Fig.~\ref{fig:IX}d that compares emission of unstrained regions of the HBL with that recorded from three pillar sites (labelled A, B and C as in Fig.~\ref{fig:IX}a). The characteristic IX emission at \SI{1.38}{\electronvolt} can still be observed in each emission spectrum since the probed sample volume is much larger than the nanopillar size. However, additional peaks emerge in the vicinity of the nanopillar sites, red-shifted by up to $\sim$\SI{100}{\milli\electronvolt} from the peak of the free IX emission. The observation of a pronounced red-shift is similar to reports of strain-induced bandgap modulation in monolayer MoSe$_2$ \cite{Branny2016}, MoS$_2$ \cite{Li2015} and particularly WSe$_2$ \cite{Branny2017,Palacios-Berraquero2017}, where this effect is routinely used to generate quantum emitters. We attribute these features to \textit{localized} interlayer excitons (LIXs) at the pillar sites, most likely due to strain-related trapping potentials around the nanopillar. The interlayer exciton character is confirmed by PL-excitation spectroscopy detected on the LIX features, that clearly reveals free exciton fingerprints of both MoSe$_2$ and WSe$_2$ component monolayers, and time-resolved PL measurements, that show similar radiative decay times for localized and free IXs (see Supplementary Notes 3 and 4). To illustrate the confinement of the LIX features to the pillar sites, the inset of Fig.~\ref{fig:IX}d shows the spatially resolved emission intensity of an example LIX peak at pillar site A. 
The spatial extent of this feature is clearly limited by the spatial resolution of our $\mu$PL setup (\SI{1}{\micro\meter}), indicative of a point-like source. Figures~\ref{fig:IX}e~and~\ref{fig:IX}f compare normalized IX $\mu$PL spectra plotted as a function of excitation power from the flat region of the HBL and on pillar A, respectively. In both cases, free IX emission is observed that blueshifts by up to $\sim \SI{20}{\milli\electronvolt}$ over the examined power range.
The spectra recorded from pillar~A additionally exhibit discrete LIX emission lines. \textit{Individually}, these features do \textit{not} blueshift with excitation power, in strong contrast to the broad IX peak. Instead, changes of excitation power result in a redistribution of the PL intensity among the different LIX peaks. However, if one analyzes the entirety of all LIX emission peaks and determines its center of the spectral weight as a function of excitation power (see Suppl.~Fig.~5), one finds a collective LIX blueshift which is slightly stronger than that of the free IX, consistent with stronger interactions for IXs spatially confined to a trapping potential.
This effect of the LIX emission shifting among several discrete peaks with varying excitation power was observed for all pillars exhibiting LIX emission and was reproducible following thermal cycling.


\subsection*{Spectroscopy of trapped few-exciton states}


\begin{figure*}
\includegraphics{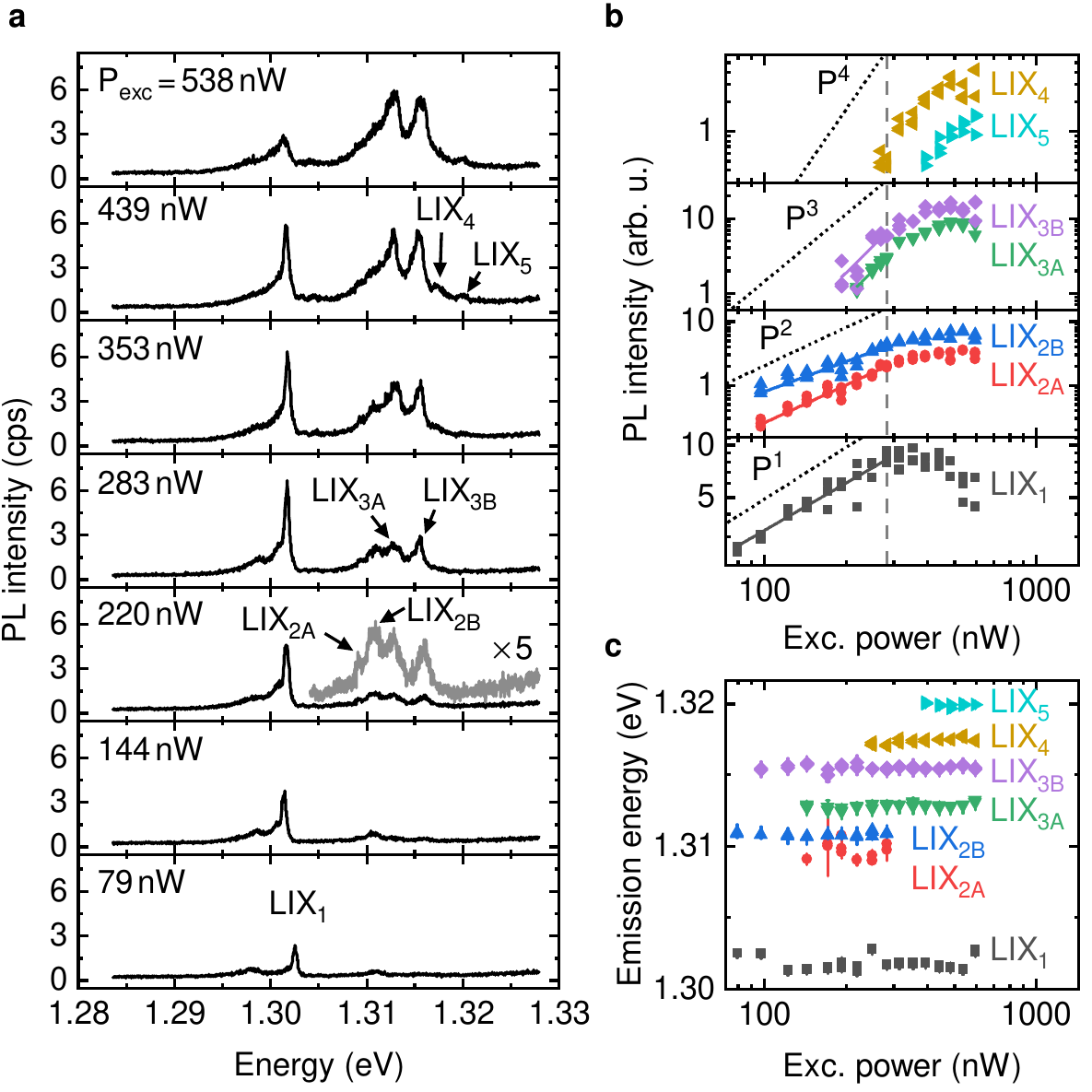}
\caption{\textbf{Few-particle interactions in a sequentially filled IX trapping potential.} \textbf{a} PL spectra of localized IX at the same pillar for varying powers. \textbf{b} Intensity of individual emission peaks determined by integrating the signal in a narrow spectral region around the peak center plotted as a function of excitation power. The characteristic power exponents of the intensity dependencies are given in the main text. The gray dashed line denotes the onset of saturation effects. \textbf{c} Emission energies of the different $LIX_i$ peaks presented in panel~a plotted as a function of excitation power. Error bars represent fit uncertainties}
\label{fig:LIX-Power}
\end{figure*}

The LIX emission spectra of pillar~A for excitation powers ranging from \SI{79}{\nano\watt} to $\geq \SI{500}{\nano\watt}$, two orders of magnitude weaker than the power used to record the results depicted in Fig.~\ref{fig:IX}, are presented in Fig.~\ref{fig:LIX-Power}a. At the lowest excitation power, the emission spectrum is found to be dominated by a single sharp line at \SI{1.302}{\electronvolt}, labelled LIX$_1$ in the figure. The spectral linewidth of this peak is \SI{550}{\micro\electronvolt} and it is broadened on the low-energy flank, most likely due to coupling to acoustic phonons\cite{Abramson2018}. The integrated intensity of LIX$_1$ is plotted in Fig.~\ref{fig:LIX-Power}b as a function of the excitation power. From the lowest excitation levels investigated it increases linearly with power, as can be seen from the solid line on the figure that shows a fit to the data with $I = I_0 \cdot P^\alpha$ yielding $\alpha = 0.9 \pm 0.1$. The dotted line on the figure shows our expectations for a linear power dependence. This observation, combined with the saturation and eventual decrease of the peak intensity at the highest excitation levels investigated, clearly identifies this peak as stemming from the recombination of a single IX in the local trapping potential, as is commonly observed in the emission characteristics of III-V quantum dots and other discrete quantum emitters\cite{Zrenner2000}.  As the excitation power is increased from the minimum of \SI{79}{\nano\watt}, additional emission features emerge at higher energies, the most prominent of which is a pair of lines labelled LIX$_{2\text{A}}$ and LIX$_{2\text{B}}$ in Fig.~\ref{fig:LIX-Power}. In contrast to LIX$_1$, LIX$_{2\text{A/B}}$ \textit{both} exhibit a near quadratic power dependence ($\alpha_{2\text{A}} = 1.96\pm 0.15$, $\alpha_{2\text{B}} = 1.57\pm 0.12$) and each saturates and quenches at the highest pump powers investigated (see Fig.~\ref{fig:LIX-Power}b). For excitation power in excess of $\sim\SI{300}{\nano\watt}$, additional manifolds of emission lines emerge, labelled LIX$_{3\text{A/B}}$, LIX$_4$ and LIX$_{5}$ on Fig.~\ref{fig:LIX-Power}a. The power dependence of each emergent set of lines clearly becomes successively more superlinear in behaviour - LIX$_{3\text{A/B}}$ have a power dependence that is very close to cubic ($\alpha_{3\text{A}} = 3.48\pm 0.27$, $\alpha_{3\text{B}} = 3.6\pm 0.7$), LIX$_4$ close to quartic (see dotted lines on figure). These observations mirror previous findings for multi-exciton states in semiconductor quantum dots\cite{Zrenner2000} and double quantum wells.\cite{Schinner2013} Since there were only a few data points before saturation for some of the peaks, we substantiate our particle number assignment by comparing ratios of intensities of two peaks to an independent capture model (see Suppl.~Fig.~6). This method uses data from the entire available power range.

Figure~\ref{fig:LIX-Power}c shows the emission energies of the various LIX peaks as extracted from multiple peak fitting. The energy of the individual LIX$_i$ peaks is not significantly power dependent, beyond spectral wandering. This can be prominently seen in the LIX$_1$ peak, which jitters by up to \SI{\sim1}{\milli\electronvolt} between measurements. Similar spectral wandering has also been reported from discrete emitters in monolayer WSe$_2$, where it is commonly attributed to time-dependent fluctuations in the local electrostatic environment. \cite{Iff2017,Srivastava2015} The mid-point of the biexciton LIX$_2$ doublet is significantly blueshifted by $\SI{8.4 \pm 0.6}{\milli\electronvolt}$ with respect to LIX$_1$ and the two components are separated by $\sim \SI{1.2\pm0.5}{\milli\electronvolt}$.


\subsection*{Modelling interactions in multi-exciton complexes}


\begin{figure}
\includegraphics{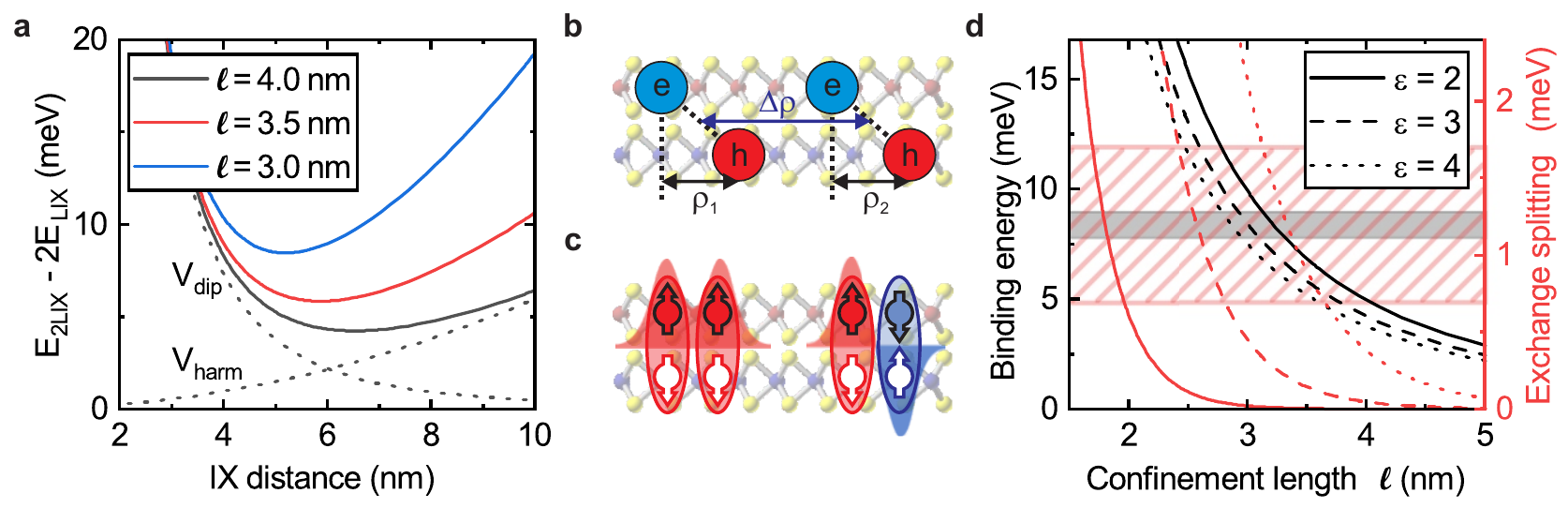}
\caption{\textbf{Theoretical model for the observed biexciton blueshift and splitting.} \textbf{a}, Total energy of two trapped IX composed of dipole-dipole repulsion and an attractive harmonic potential (dashed lines) for different confinement lengths. We assume $\varepsilon_r=3$ and $m_{e/h} = 0.8/0.45 m_0$ \textbf{b}, Sketch of the interlayer biexciton complex and the coordinate conventions used in the main text. \textbf{c}, Schematic illustration of biexciton states with different symmetries of spatial and spin wavefunction. \textbf{d}, Binding energy (gray curves) and exchange splitting (red curves) of the trapped biexciton as a function of the confinement length of the harmonic potential. Positive binding energies correspond to an increase in energy. The gray-shaded (red-shaded) energy range indicates the measured biexciton blueshift (splitting).}
\label{fig:LIX2-Model}
\end{figure}

In order to quantitatively understand the strong blueshift of the localized interlayer biexciton from the single exciton and the splitting of the biexciton into a doublet, we modelled the interaction of two IX confined to a harmonic trapping potential, as characterised by a confinement lengthscale $\ell=\sqrt{\hbar/m^*\Omega_0}$. Here, $m^*$ denotes the total effective mass of the exciton and $\Omega_0$ is the angular frequency of an exciton in the harmonic potential. Figure~\ref{fig:LIX2-Model}a illustrates how we calculate the equilibrium separation $\Delta\rho$ between two IX by minimizing the sum of repulsive (dipolar) and attractive (harmonic) potential energies. For values of $\ell\sim\SIrange{3}{4}{\nano\meter}$, reference to Fig.~\ref{fig:LIX2-Model}a shows that $\Delta\rho\sim\SIrange{5}{6.5}{\nano\meter}$. We now continue to discuss the calculation of the direct and exchange Coulomb interaction energies for the $LIX_2$ and $LIX_1$ states. For this purpose, we modified the method outlined in Ref.~\cite{Bondarev2018} and start with a Hamiltonian describing the biexciton complex of the following form:
\begin{equation}\label{eq:1}
H = H_0 + H_\mathrm{int},
\end{equation}
where
\begin{equation}\label{eq:2}
H_0 = \sum_{i=1,2} \left [ -\frac{\hbar^2}{2m^*} \frac{1}{\rho_i} \frac{\partial}{\partial \rho_i} \left ( \rho_i \frac{\partial}{\partial \rho_i} \right ) - \frac{1}{4\pi\varepsilon} \left ( \frac{1}{\sqrt{\rho_i^2 + d^2}} + \frac{1}{\sqrt{(\rho_i + (-)^i \Delta \rho)^2 + d^2}} \right ) + \frac{1}{2} m^* \Omega_0^2 \rho_i^2 \right ]
\end{equation}
describes non-interacting excitons exposed to a weak harmonic confinement potential and
\begin{eqnarray}\label{eq:3}
H_\mathrm{int} & & = \frac{e^2}{\pi\varepsilon} \Bigg [ \frac{1}{\left | \rho_1-\rho_2 + 2\Delta\rho \right |} + \frac{1}{\left | \rho_1-\rho_2 - 2\Delta\rho \right |} \\
& & - \frac{1}{\sqrt{\left ( \rho_1 + \rho_2 + 2\Delta \rho \right )^2 + d^2}} - \frac{1}{\sqrt{\left ( \rho_1 + \rho_2 - 2\Delta \rho \right )^2 + d^2}} \Bigg] \nonumber
\end{eqnarray}
represents the Coulomb interactions between the excitons. In equations \ref{eq:2} and \ref{eq:3}, the coordinates $\rho_1$ and $\rho_2$ are in-plane projections of the relative electron-hole coordinates of the interlayer excitons, as depicted in Fig.~\ref{fig:LIX2-Model}b, and $\Delta \rho$ was introduced before. Moreover, $d \approx 0.7~$nm is the interlayer separation of the HBL that determines the strength of the dipole-dipole repulsion and $\epsilon=\epsilon_0\epsilon_r$ describes the effective dielectric screening by the surrounding vacuum (upper) and SiO$_2$ (lower) media. We note that, in the simplest picture in which we average the values for vacuum (upper cladding) and SiO$_2$ (lower substrate), the effective dielectric constant would be $\epsilon_r=(\epsilon_{vac}+\epsilon_{SiO_2})/2\sim2.5$. For the effective masses $m^*$, we use $m^*=0.8 m_0$ ($0.45m_0$) for the MoSe$_2$~electrons\cite{Larentis2018} (WSe$_2$~holes\cite{Fallahazad2016}). 

The indistinguishability of the two LIX composite bosons forming the biexciton implies that the total wavefunction must be symmetric with respect to IX exchange. This condition can only be satisfied when both spin and spatial parts of the two-exciton wavefunction are simultaneously \textit{symmetric} or \textit{antisymmetric}, respectively, as illustrated schematically in Fig.~\ref{fig:LIX2-Model}c. Thus, the spatial part of the two-IX (biexciton) wavefunction can take the form $\Psi\sim\frac{1}{\sqrt{2}}[\psi_{IX}(\rho_1)\psi_{IX}(\rho_2)\pm\psi_{IX}(\rho_2)\psi_{IX}(\rho_1)]$, giving rise to two, energetically distinct biexciton states separated by the exchange energy $2J(\Delta\rho)$. Here, $J(\Delta\rho)$ is given by \cite{Bondarev2011}
\begin{equation}\label{eq:4}
J(\Delta\rho) = \frac{1}{3} \int_{-\Delta \rho/\sqrt{2}}^{\Delta \rho/\sqrt{2}} \mathrm{d} y \big | \Psi(x,y) \frac{\partial \Psi(x,y)}{\partial x} \big |_{x=0},
\end{equation}
where $\Psi(x,y)$ is the ground state of eqn. \eqref{eq:1} that has been transformed to the coordinate system $x = (\rho_1 - \rho_2 - \Delta\rho)/\sqrt{2}$ and $y = (\rho_1 + \rho_2)/\sqrt{2}$.  From this expression, we numerically calculate the exchange splitting to obtain the absolute energies of the two biexciton states:
\begin{equation}\label{eq:5}
E_\mathrm{LIX2B/A} = 2 E_\mathrm{LIX1} + \frac{2d^2}{4\pi\varepsilon (\Delta \rho)^3} + \frac{\hbar ^2 \left( \Delta \rho \right)^2}{4m^* l^4}  \pm J \left( \Delta \rho \right) = 2 E_\mathrm{LIX1} + E_\mathrm{binding}  \pm J \left( \Delta \rho \right)
\end{equation}

Here, a positive binding energy refers to an increase in energy of the biexciton state compared to two isolated excitons. Representative results of our calculations are presented in Fig.~\ref{fig:LIX2-Model}d that shows the numerically evaluated biexciton binding energy and exchange splitting as a function of the effective confinement lengthscale $\ell$. The three curves presented on the figure correspond to effective dielectric constants $\epsilon=2-4$. The experimentally observed blueshift of LIX$_2$ of \SI{8.4 \pm 0.6}{\milli\electronvolt} is consistent with a confinement lengthscale in the range $\ell=\SIrange{2.8}{3.3}{\nano\meter}$, with relatively weak sensitivity to the effective dielectric constant. We interpret the existence of two biexciton states, split by \SI{1.2\pm0.5}{\milli\electronvolt}, as reflecting the existence of two distinct spatial wavefunctions with even and odd symmetry with respect to IX exchange, as discussed above. The red curves presented in Fig.~\ref{fig:LIX2-Model}d show the calculated exchange splittings $J(\ell)$ for the same range of effective dielectric constants. The red-shaded region indicates the measured splitting of the biexciton peak. Remarkably, our model calculations simultaneously reproduce \textit{both} the observed biexciton binding energy and the exchange splitting for a dielectric constant in the range $\epsilon \gtrapprox 3$, very close to the average dielectric constant of the environment $\epsilon_{vac-SiO_2}=2.5$, and an effective confinement lengthscale of $\ell\approx \SI{3}{\nano\meter}$.  Thus, we conclude that the strain-induced trapping potentials in the vicinity of the nanopillars give rise to nanometer-scale confinement traps in which the LIX interact via direct and exchange Coulomb interactions.

Figure~\ref{fig:LIX-Power}c also shows that the triexciton (LIX$_3$), quadexciton (LIX$_4$) and quintexciton (LIX$_5$) transition lines are even more strongly blueshifted from LIX$_1$, as compared to LIX$_2$. We measure blueshifts of \SI{12.4\pm0.4}{\milli\electronvolt} for LIX$_3$, \SI{15.5\pm0.6}{\milli\electronvolt} for LIX$_4$ and \SI{18.2\pm0.8}{\milli\electronvolt} for LIX$_5$. We extended our calculations for binding energies based on dipole-dipole repulsion in the harmonic trap to three to five excitons (see Supplementary Note 10). The total energy is minimized by maximizing the equilibrium spacing between the particles. Thus, the IXs in LIX$_3$ (LIX$_4$/LIX$_5$) would be expected to be located at the corners of an equilateral triangle (square/regular pentagon) centered around the trap minimum. We find that independent of the system parameters (e.g. dipole moment $d$ or confinement length $\ell$) the blueshift of the LIX$_3$ (LIX$_4$/LIX$_5$) is expected to be about $1.36\times\Delta E_{\mathrm{LIX2}}$ ($1.92\times\Delta E_{\mathrm{LIX2}}$/$2.53\times\Delta E_{\mathrm{LIX2}}$), where $\Delta E_{\mathrm{LIX2}}$ is the localized biexciton blueshift. Remarkably, the experimentally observed shifts amount to $\left(1.48 \pm 0.12 \right)\times\Delta E_{\mathrm{LIX2}}$ ($\left( 1.85 \pm 0.15 \right) \times\Delta E_{\mathrm{LIX2}}$, $\left(2.17 \pm 0.18 \right)\times\Delta E_{\mathrm{LIX2}}$), respectively, in excellent agreement with this simple prediction.


In summary, we have directly probed interactions between a discrete number $N_{IX}=\numrange{1}{5}$ of interlayer excitons moving within a localized trapping potential. Power-dependent measurements allow us to identify different few-body states and, moreover, the observed interactions are found to be all mutually in agreement with a localization lengthscale $\ell \approx \SI{3}{\nano\meter}$. We reproduce the observed blueshift of the biexciton by \SI{8.4 \pm 0.6}{\milli\electronvolt} as well as the observed biexciton fine structure with a splitting of \SI{1.2\pm0.5}{\milli\electronvolt} in a model including the dipolar and exchange interaction of two IX in a harmonic trapping potential. The relative energies of states with one to five trapped IXs are well captured by our model, which predicts maximal spacing of the IXs within the limitations of the trapping potential. The significant interaction energies observed in this paper suggest that strongly correlated multiexciton phenomena, such as Wigner crystallization, may be realized in layered van der Waals heterostructures, opening up new avenues for nonlinear coherent optical control and spin-optronics with interlayer excitons.

After preparation and submission of the manuscript, we became aware of similar results by another group. \cite{Li2019}


\section{Methods}
\textit{Sample fabrication:} We fabricated the sample on a Si-[100] wafer with a \SI{285}{\nano\meter} thick SiO$_2$ layer on top. We defined the pillar grid via electron beam lithography on a negative resist (AR-N 7520.07) and etched it into the substrate using F-based reactive ion etching. The pillars have a diameter and height of $\sim$\SI{130}{\nano\meter} and $\sim$\SI{90}{\nano\meter}, respectively. We exfoliated crystals of WSe$_2$ and MoSe$_2$ (hqgraphene.com) onto polydimethylsiloxane (PDMS, Gel-Pak) stamps, identified monolayers in an optical microscope via reflectivity contrast and then subsequently transferred the WSe$_2$ and MoSe$_2$ using a mask aligner in a cleanroom environment. Each transfer was followed by an annealing step in vacuum (\SI{200}{\celsius}, \SI{30}{\minute} after the WSe$_2$ transfer; \SI{150}{\celsius}, \SI{20}{\minute} after the MoSe$_2$ transfer which completed the heterostructure). During assembly of the HBL, we took efforts to orientate the crystal axes of the two materials by aligning long crystalline edges of the monolayers.

\textit{Optical spectroscopy measurements:} Photoluminescence measurements were carried out in two different setups designed for low-temperature confocal microscopy. The data from Fig.~\ref{fig:IX} was acquired at a setup using a liquid-He flow cryostat at $T = \SI{10}{\kelvin}$ and a HeNe laser ($\lambda = \SI{633}{\nano\meter}$) focused to a diameter of $\sim \SI{1}{\micro\meter}$. In order to counteract thermal cycling effects (see Supplementary Note 11), the measurements presented in Fig.~\ref{fig:LIX-Power} were carried out in a setup equipped with a closed-cycle cryostat at $T=\SI{4}{\kelvin}$. The excitation laser used in this setup was at $\lambda = \SI{532}{\nano\meter}$ and was defocused due to chromatic aberration in the used objective.

We note that the data presented in Fig.~\ref{fig:IX}f~and~\ref{fig:LIX-Power} were both recorded from pillar A, but following thermal cycling of the sample to room temperature. This procedure was typically found to result in a shift of the absolute energy of different groups of sharp-line emission features by $\sim\SI{\pm10}{\milli\electronvolt}$. Nevertheless, following each cool-down, the generic form of the group of peaks was retained. Supplementary~Figure~5 shows power-dependent $\mu PL$ spectra recorded from pillar~A (from the same measurement as the data presented in Fig.~\ref{fig:IX}f), illustrating how the single sharp emission feature is observed close to \SI{1.292}{\electronvolt} at the lowest excitation power investigated, with additional peaks emerging in the range $\sim \SIrange{5}{30}{\milli\electronvolt}$ to higher energy as $P_{ex}$ increases.

\textit{Data analysis:} The emission energies shown in Fig.~\ref{fig:LIX-Power}c were extracted from Gaussian fits to the PL spectra. In some power regimes, the large peak overlap combined with the low signal-to-noise ratio hindered proper fitting of the data. For this reason, the PL intensities presented in Fig.~\ref{fig:LIX-Power}b were determined by integrating the PL signal in an interval of \SI{0.12}{\milli\electronvolt} around the respective center energies. The only exception is the LIX$_1$ peak for which fitting was possible over the whole power range due to its spectral isolation. The data points for LIX$_{3\mathrm{A}}$ and LIX$_{3\mathrm{B}}$ in Fig.~\ref{fig:LIX-Power}b initially overlapped. Thus, the data for LIX$_{3\mathrm{A}}$ was slightly offset for better visibility.

\section{Data availability}
The data that support the findings of this study are available from the corresponding author upon reasonable request.

\section{Acknowledgements}
We gratefully acknowledge funding from the International Max Planck Research School for Quantum Science and Technology (IMPRS-QST), the DFG Clusters of Excellence \mbox{MCQST}, e-Conversion and NIM, DFG Project FI 947/5-1 SQAM, the European Union's Horizon 2020 research and innovation programme under grant agreement No. 820423 (S2QUIP), the BMBF \textit{Verbundprojekt} QLink-X and the Alexander von Humboldt Foundation. K.M. acknowledges support from the Bavarian Academy of Sciences and Humanities. In addition, we thank Richard Schmidt and his team for very helpful discussions.

\section{Author contributions}
M.K., B.D.G., K.M. and J.J.F. conceived and designed the experiments. M.K. and J.G. prepared the sample. M.K., M.B.-G., J.G. and M.M. performed the experiments. M.K. analyzed the data. J.K. modelled the exchange interaction. M.K., K.M. and J.J.F. wrote the manuscript with input from all authors.

\section{Competing interests}
The authors declare that there are no competing interests.


\providecommand{\latin}[1]{#1}
\makeatletter
\providecommand{\doi}
  {\begingroup\let\do\@makeother\dospecials
  \catcode`\{=1 \catcode`\}=2 \doi@aux}
\providecommand{\doi@aux}[1]{\endgroup\texttt{#1}}
\makeatother
\providecommand*\mcitethebibliography{\thebibliography}
\csname @ifundefined\endcsname{endmcitethebibliography}
  {\let\endmcitethebibliography\endthebibliography}{}

\clearpage

\setcounter{equation}{0}
\setcounter{table}{0}
\renewcommand{\tablename}{Supplementary Table}
\setcounter{figure}{0}
\renewcommand{\figurename}{Supplementary Figure}
\setcounter{section}{0}
\setcounter{subsection}{0}
\setcounter{page}{1}

\section*{Supplementary Information to: \textit{Discrete Interactions between a few Interlayer Excitons Trapped at a MoSe$_2$-WSe$_2$~Heterointerface}}


\subsection*{Supplementary Note 1: Enhanced micrograph of the sample}
\label{sub:SampleZoom}

\begin{figure*}[h]
\includegraphics{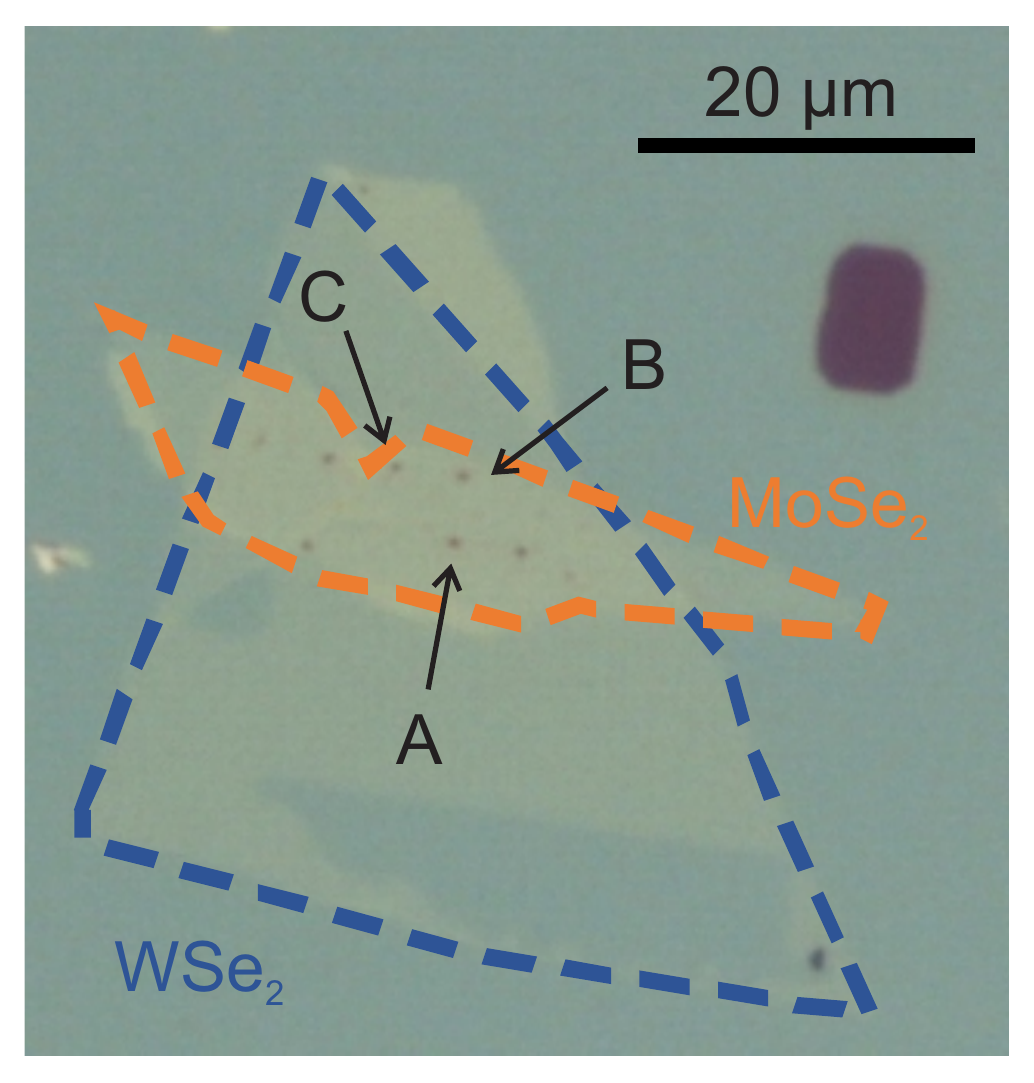}
\caption{\textbf{Optical micrograph, magnified in comparison to Fig.~1b.} The monolayers of WSe$_2$ and MoSe$_2$ are outlined with blue and orange dashed lines, respectively. The nanopillars appear as black dots in the heterobilayer region. Nanopillars, from which PL data are presented, are labelled accordingly.}
\label{fig:SampleZoom}
\end{figure*}
\newpage


\subsection*{Supplementary Note 2: Excitation-power dependence of free IX emission}
\label{sub:FreeIX}

\begin{figure*}[h]
\includegraphics{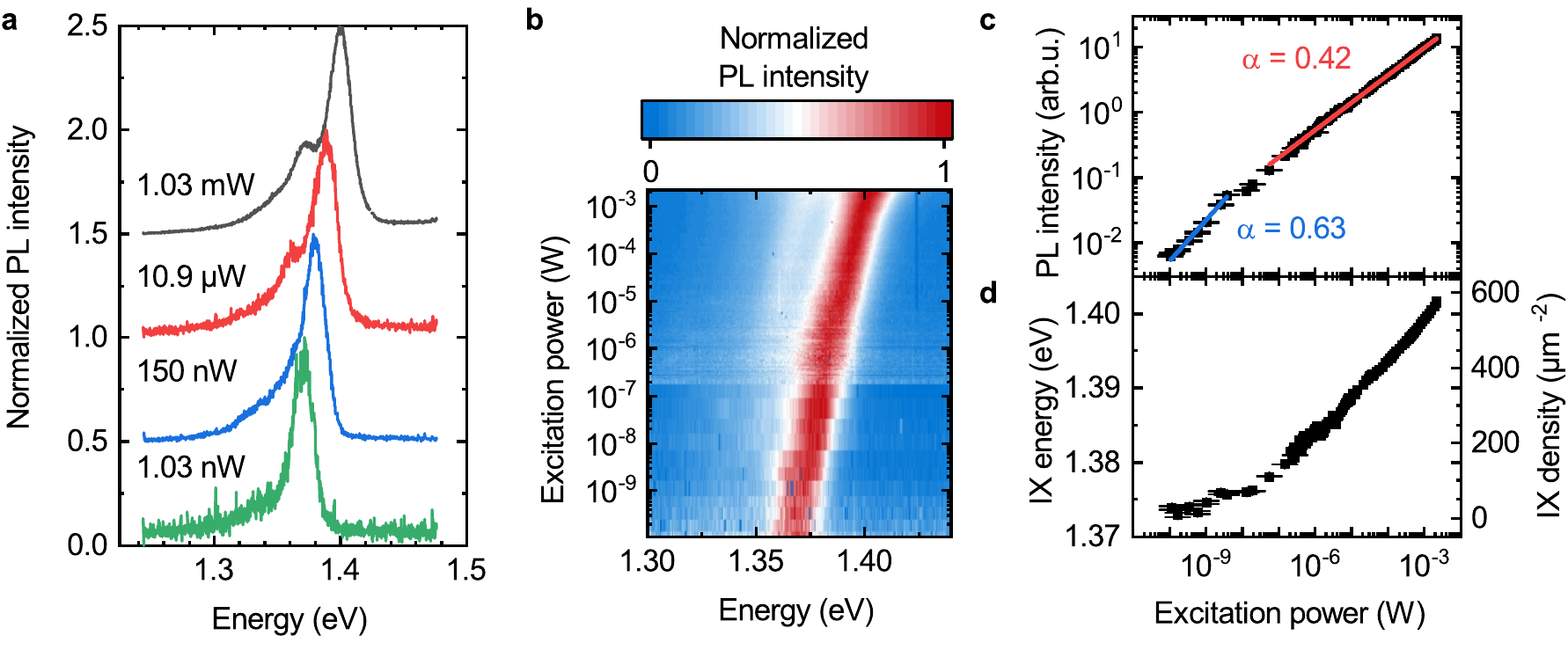}
\caption{\textbf{Excitation-power-dependent emission from IX in the unstrained HBL region.} \textbf{a} Normalized example IX spectra at different excitation powers. \textbf{b} False color plot of normalized IX spectra as a function of excitation power. \textbf{c} PL intensity of the main IX peak as function of excitation power. The intensity was determined from Gaussian fits to the data. The solid lines indicate power law fits of the form $I=I_0 \cdot P^\alpha$ to the data. \textbf{d} Emission energy of the IX main peak as a function of excitation power. The energies were determined from Gaussian fits to the data. An estimate of the IX density from dipole-dipole repulsion energies is given on the right y-axis.}
\label{fig:IXPower}
\end{figure*}

Supplementary Figure~\ref{fig:IXPower}a shows exemplary PL spectra for different excitation powers recorded from a HBL region of the sample away from SiO$_2$ nanopillars, together with comprehensive power-dependent data shown as a false-color plot in Suppl.~Fig.~\ref{fig:IXPower}b.  The power-dependence of the integrated IX intensity is plotted in Suppl.~Fig.~\ref{fig:IXPower}c on a double-logarithmic representation.  We observe a typical $\sqrt{P_{ex}}$ dependence with a power factor $\alpha =0.43\pm0.05$ (red line) consistent with exciton-exciton annihilation at elevated excitation levels\cite{Amani2015x}. Careful examination of fig.~\ref{fig:IXPower}c for $P_{ex} \leq P_{ex}^0 \approx \SI{20}{\nano\watt}$ shows a higher power factor. A power factor of $\alpha = 0.66$ would be consistent with Auger recombination. The increased power factor could also hint at a transition towards a regime dominate by radiative recombination ($\alpha = 1$). An unambiguous trend cannot be extracted from the data. Supplementary~Figure~\ref{fig:IXPower}d shows the peak energy of the IX emission as determined from Gaussian fits, illustrating a strong $P_{ex}$-dependent blueshift beyond the threshold power $P_{ex}^0$, most likely reflecting the dipolar exciton-exciton interactions.  For $P_{ex}\leq P_{ex}^0$ a much weaker blueshift is observed in our experiments. The simplest analysis of the blueshift based on a "plate capacitor formula" would indicate that $\Delta E=(4\pi n e^2 d/\kappa)$\cite{Butov1999x}, where $n$ is the areal concentration of IX, $d$ is the effective separation of charge along the z-axis and $\kappa$ the effective dielectric constant. The righthand axis of Suppl.~Fig.~\ref{fig:IXPower}d shows the mapping of the observed blueshift onto the areal IX density using $d=\SI{0.7}{\nano\meter}$ and $\kappa=4$. In more realistic models the effects of spatial inhomogenities in the IX density, finite temperature effects and multi-particle correlations must be properly taken into account\cite{Laikhtman2009x}.

\newpage


\subsection*{Supplementary Note 3: Photoluminescence excitation spectroscopy}
\label{sub:PLE}

\begin{figure*}[h]
\includegraphics{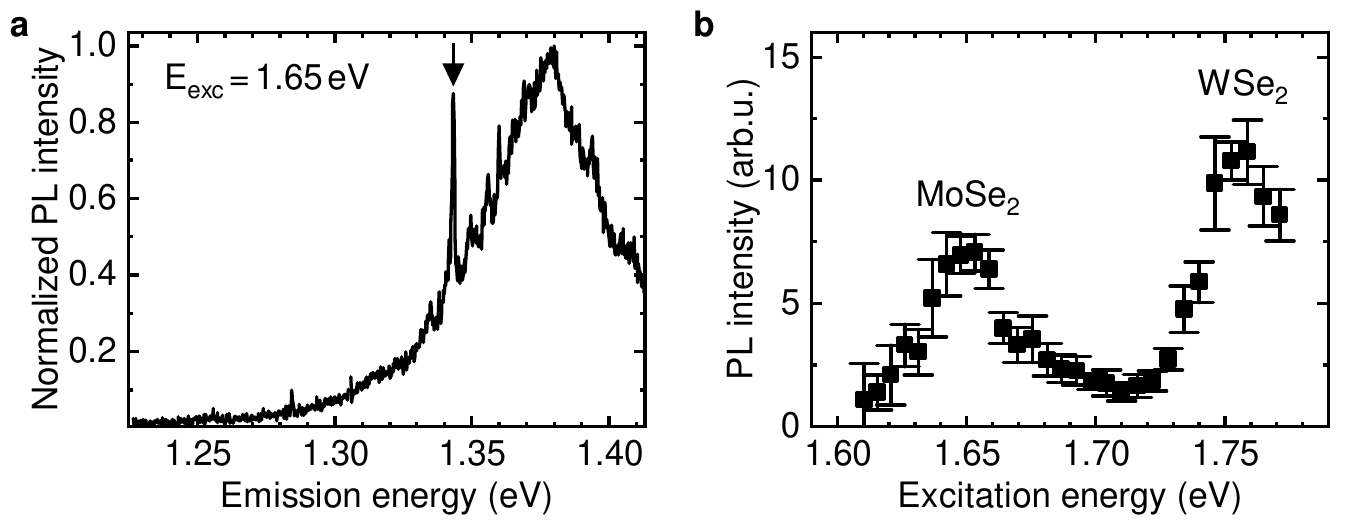}
\caption{\textbf{Photoluminescence excitation spectroscopy of localized interlayer excitons.} \textbf{a} Example low-power PL spectrum at pillar A with an excitation energy of \SI{1.65}{\electronvolt}. \textbf{b} Intensity of the LIX peak marked in panel~a as a function of excitation energy.}
\label{fig:PLE}
\end{figure*}

We recorded the PL intensity of an example LIX peak (marked in Suppl.~Fig.~\ref{fig:PLE}a) as a function of excitation energy. This dependence is shown in Suppl.~Fig.~\ref{fig:PLE}b and exhibits two clear peaks in the spectral range of MoSe$_2$ (\SI{1.65}{\electronvolt}) and WSe$_2$ (\SI{1.75}{\electronvolt}) excitons. This behaviour was also observed for other LIX peaks, including peaks identified as multi-IX at higher excitation powers, and is consistent with free IXs\cite{Rivera2015x,Nagler2017x,Seyler2018x}.

\newpage


\subsection*{Supplementary Note 4: Time-resolved photoluminescence}
\label{sub:TRPL}

\begin{figure*}[h!]
\includegraphics{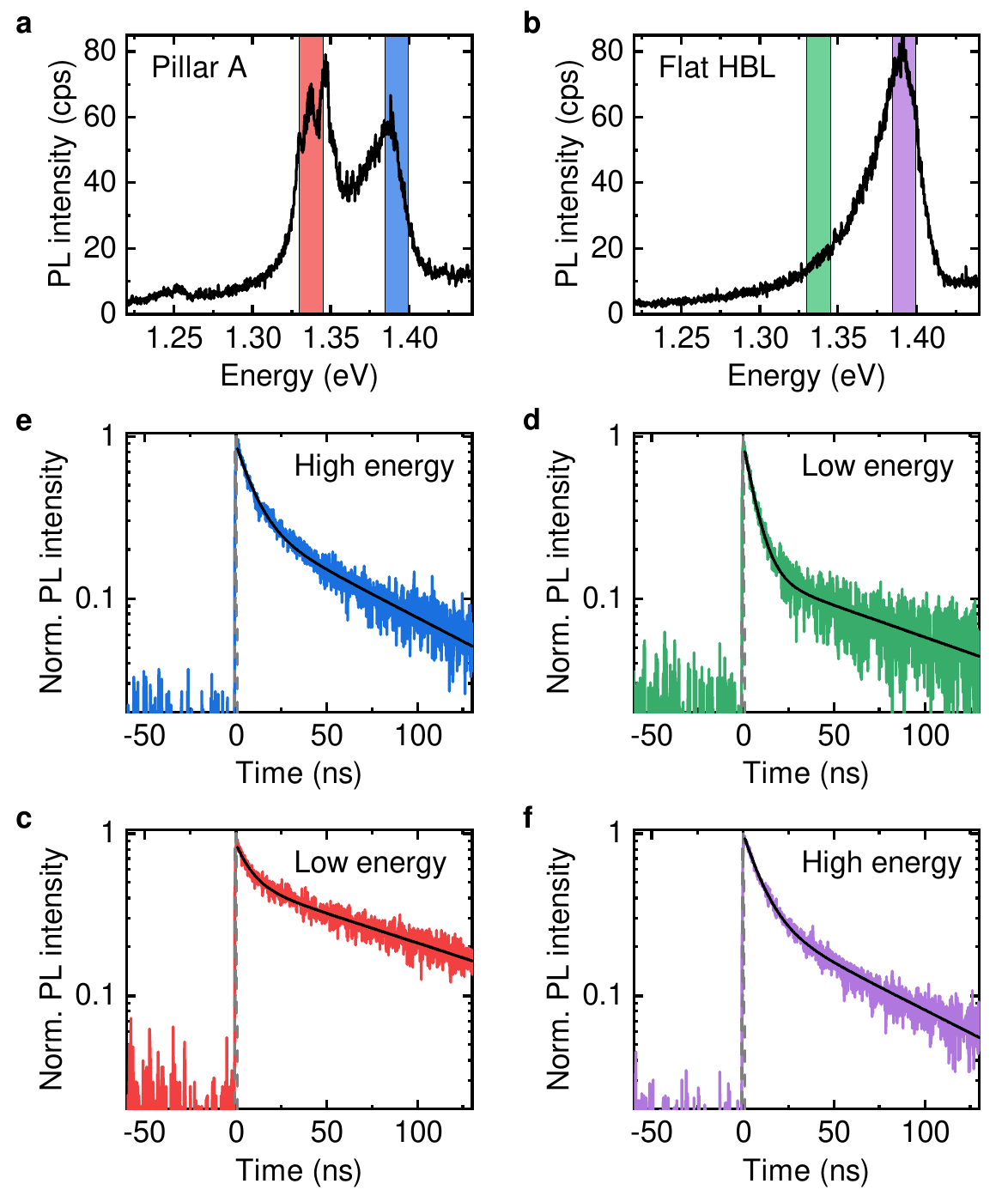}
\caption{\textbf{Time-resolved photoluminescence of free and localized interlayer excitons.} \textbf{a,b} PL spectra at pillar A (a) and a flat HBL reference position (b) under pulsed excitation (\SI{90}{\pico\second} pulses at \SI{640}{\nano\meter} with a pulse energy of \SI{17.2}{\pico\joule} and \SI{2.5}{\mega\hertz} repition rate). \textbf{c-f} Time-resolved photoluminescence at different sample positions and spectral ranges. The curves' colors match the spectral indicators in panels~a~and~b. The gray dashed lines denote the instrument response and the black solid lines show biexponential fits to the data.}
\label{fig:TRPL}
\end{figure*}

\begin{table}[h]
    \centering
    \begin{tabular}{c c | c c c c}
        \hline \hline
        Position & Spectral range & Slow timescale & Slow proportion & Fast timescale & Fast proportion \\
       \hline
       Pillar & Low energy & \SI{118.8 \pm 2.1}{\nano\second} & $(95 \pm 3) \%$ & \SI{8.1 \pm 0.5}{\nano\second} & $(5 \pm 3) \%$\\
       Flat HBL & Low energy & \SI{111\pm 4}{\nano\second} & $(76\pm 5)\%$ & \SI{6.16\pm 0.12}{\nano\second} & $(24\pm 5)\%$\\
       Pillar & High energy & \SI{74.5\pm 1.5}{\nano\second} & $(80\pm 3)\%$ & \SI{9.0\pm 0.3}{\nano\second} & $(20\pm 3)\%$\\
       Flat HBL & High energy & \SI{76.5\pm 1.4}{\nano\second} & $(78\pm 3)\%$ & \SI{9.48\pm 0.16}{\nano\second} & $(22\pm 3)\%$\\
       \hline \hline
    \end{tabular}
    \caption{Parameters extracted from biexponential fits to the time-resolved PL data presented in fig.~\ref{fig:TRPL}. The fitted model $I(t)=A_{slow}e^{-t/\tau _{slow}}+A_{fast}e^{-t/\tau _{fast}}$ directly includes the slow/fast timescale $\tau_{slow/fast}$. The slow/fast proportion is calculated as $\frac{A_{slow/fast} \cdot \tau_{slow/fast}}{A_{slow} \cdot \tau_{slow}+A_{fast} \cdot \tau_{fast}}$.}
    \label{tab:TRPL}
\end{table}

The single- to few-exciton states discussed in detail in the main manuscript yielded only small PL signals on top of a non-neglible background from the the broad IX emission. It was thus not possible to conduct conclusive photon auto- or cross-correlation or time-resolved PL measurements on these states. In order to obtain a decent signal-to-background ratio of localized IX emission for time-resolved measurements, we instead performed experiments at higher excitation powers: We illuminated the sample with a PicoQuant LDH-P-C-640B laser diode driven by a PicoQuant PDL 800-B driver (wavelength: \SI{640}{\nano\meter}; Pulse width $< \SI{90}{\pico\second}$; Repitition rate: \SI{2.5}{\mega\hertz}) with a cw power equivalent of \SI{43}{\micro\watt} (power per pulse: \SI{17.2}{\pico\joule}). The resulting PL spectra on pillar A and a reference position in the flat HBL region are given in Suppl.~Fig.~\ref{fig:TRPL}a~and~b, respectively. At the pillar position, broad LIX emission, composed of several overlapping peaks, can clearly be seen on top of the free IX background. We cannot identify distinct states (and in particular exciton numbers) at these excitation powers anymore, but we still attribute this emission to IX trapped in the pillar-defined potential due to its spatially confined nature and the reduced energy.

We use bandpassfilters (Thorlabs FB930-10 / FB890-10; Center wavelength: \SI{930}{\nano\meter} / \SI{890}{\nano\meter}; Transmission FWHM: \SI{10}{\nano\meter}) to select emission in the spectral range of the LIX or the free IX main peak. The transmission windows are indicated by the colored bars in Suppl.~Fig.~\ref{fig:TRPL}a~and~b. We record the time-resolved PL by measuring the time difference between signal events as recorded by a SPAD and the laser trigger with a PicoHarp~300. Supplementary~Figures~\ref{fig:TRPL}c-f show the time-resolved PL measured at the two different positions and spectral ranges. Each of the panels includes the instrument response as a gray dashed line and a fit to the data with a biexponential decay ($I(t)=A_{slow}e^{-t/\tau _{slow}}+A_{fast}e^{-t/\tau _{fast}}$) as a solid black line. The color codes of the data match the indicators in panel~a~and~b. We give a summary of the derived parameters in Supplementary~Table~\ref{tab:TRPL}. The free IX PL recorded at the pillar position (with a detection volume far greater than the strain-defined potential) and at the reference position (panel~e~and~f, respectively) show the same time dependence within the error intervals obtained from the fit.

The time-resolved PL response in the low-energy (LIX) regime looks very different on the pillar and on the reference position (panels~c~and~d) at first glance; the fits to the data reveal, however, that the timescales of the biexponential decay are very similar. The reason for the different appearance of the curves in panels~c~and~d is that the fast decay accounts for 24\% of the emission on the flat HBL, whereas this value is only 5\% at the pillar site. The general presence of a fast and a slowly decaying component in all four cases may point towards exciton-density-dependent, nonradiative decay channels, as also indicated by the sublinear power-dependence of the free IX intensity (see Suppl.~Fig.~\ref{fig:IXPower}). A definite conclusion about decay mechanisms would require additional measurements, such as time-resolved PL at different excitation powers. However, the very similar timescales support the attribution of the LIX peaks as stemming from interlayer excitons.

\newpage


\subsection*{Supplementary Note 5: Comparison of interaction-induced blueshifts of free and localized IX}
\label{sub:PowerComparison}

\begin{figure*}[h!]
\includegraphics{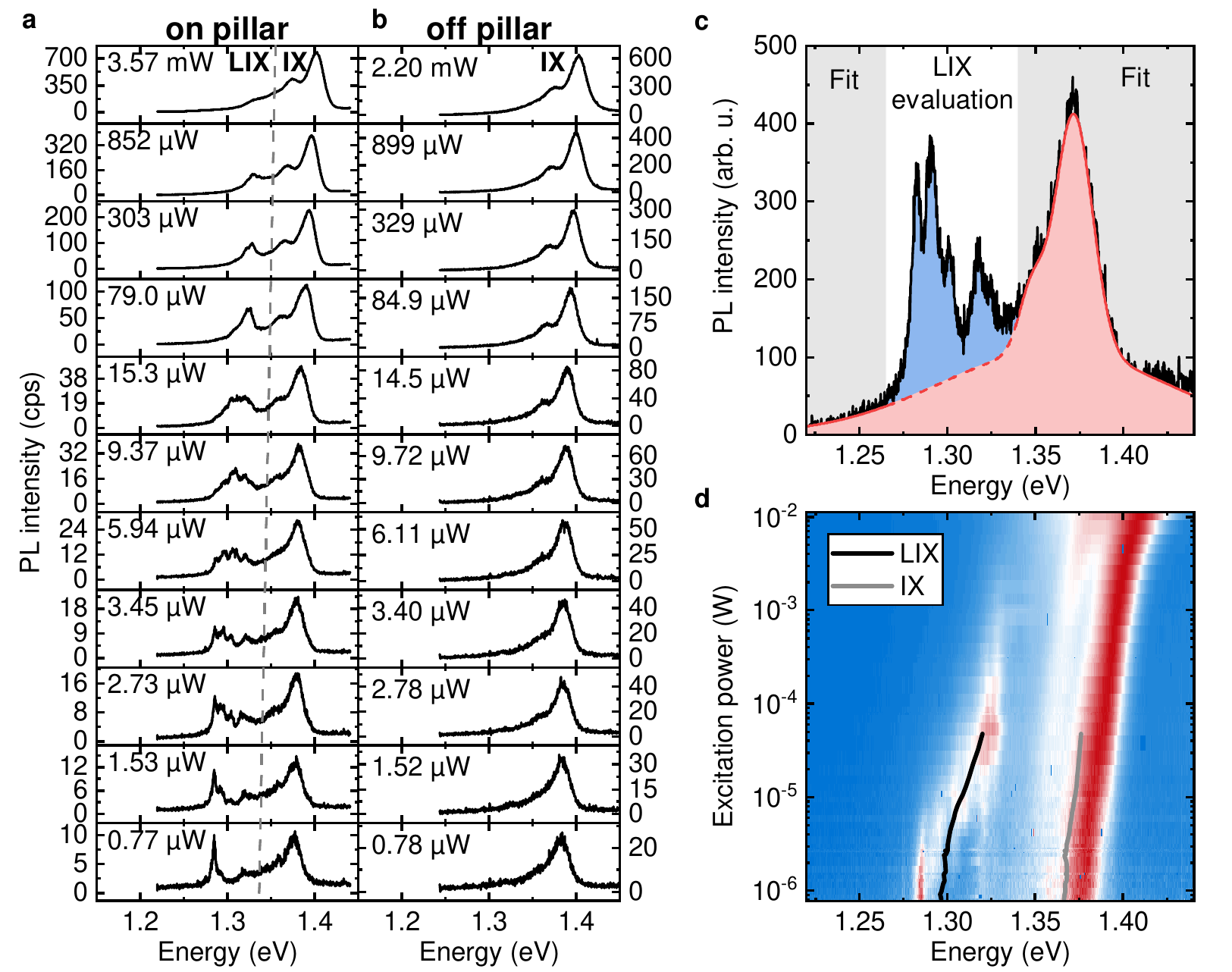}
\caption{\textbf{Excitation-power-dependet shift of IX and LIX PL emission.} All spectra in this figure were recorded at $T=\SI{10}{\kelvin}$ under excitation with a HeNe laser at \SI{633}{\nano\meter} focused to a spot of $\sim \SI{1}{\micro\meter}$ diameter. \textbf{a} Example PL spectra at varying excitation powers recorded on pillar A. \textbf{b} Example PL spectra at similar powers as presented in panel~a recorded on a reference position on the unstrained HBL. \textbf{c} Example emission spectrum at pillar A to illustrate the analysis of the spectral center of weight. The spectrum inside the fit range (indicated by the gray background) was fitted by multiple Gaussian peaks. The parameters of this fit (indicated by the red line and area) were used to determine the spectral center of emission for the free IXs. The emission inside the remaining range (labelled 'LIX evaluation') that was not accounted for by the fit (indicated by the blue area) was used to determine the spectral center of emission for the LIXs. \textbf{d} False-color plot of normalized emission spectra at different excitation powers. The black (gray) line denotes the calculated center of emission of the localized (free) IXs. While both show a trend to higher energies for higher excitation powers, this is more pronounced for the LIXs.}
\label{fig:PowComp}
\end{figure*}

Supplementary~Figures~\ref{fig:PowComp}a and \ref{fig:PowComp}b show example spectra from the datasets that are presented in Fig.~1f and 1e, respectively. Supplementary~Figure~\ref{fig:PowComp}a displays example spectra at pillar A for different powers spanning more than three orders of magnitude. As for the dataset presented in Fig.~2, we observe a series of emission features in the window \SIrange{1.28}{1.35}{\electronvolt}. As the excitation power increases from the lowest levels, the spectrum clearly evolves from a single sharp emission line to a multiplet of excitonic features. Since the overlap of the peaks in these measurements are enhanced due to the broader linewidths compared to the data presented in Fig.~2, a thorough quantitative analysis of the individual peaks was not performed for this measurement. Supplementary~Figure~\ref{fig:PowComp}b shows PL spectra recorded at a reference position inside the unstrained HBL region at comparable excitation levels. Clearly, the free IX emission shifts in a similar way to the LIX emission band in the limit of strong excitation. The doublet structure of the IX emission, clearly identified at the highest excitation levels consists of two emission features separated by $\sim \SI{25}{\milli\electronvolt}$, most likely arises from the spin-orbit split conduction band minima in the MoSe$_2$.

In order to quantitatively compare the blueshifts of LIXs and free IXs, we fit the free IX emission at a pillar site outside the LIX spectral range using multiple Gaussians. This fit is used to estimate the free IX contribution inside the LIX spectral range, as illustrated in Suppl.~Fig.~\ref{fig:PowComp}c. We calculated the spectral "center of mass" of the fitted IX and of the LIX (everything that was not accounted for by the fit). The calculated centers of mass over almost two orders of magnitude in power are superimposed onto a false-color plot of the excitation-power-dependent PL spectra in Suppl.~Fig.~\ref{fig:PowComp}d. It can be seen that both show a similarly increasing trend but the blueshift is more pronounced for the LIXs. This is expected since the dipolar interactions between IXs should be enhanced by the confinement inside a trapping potential.

\newpage


\subsection*{Supplementary Note 6: Excitation-power dependence of LIX emission at additional pillar sites}
\label{sub:OtherPillars}

\begin{figure*}[h!]
\includegraphics{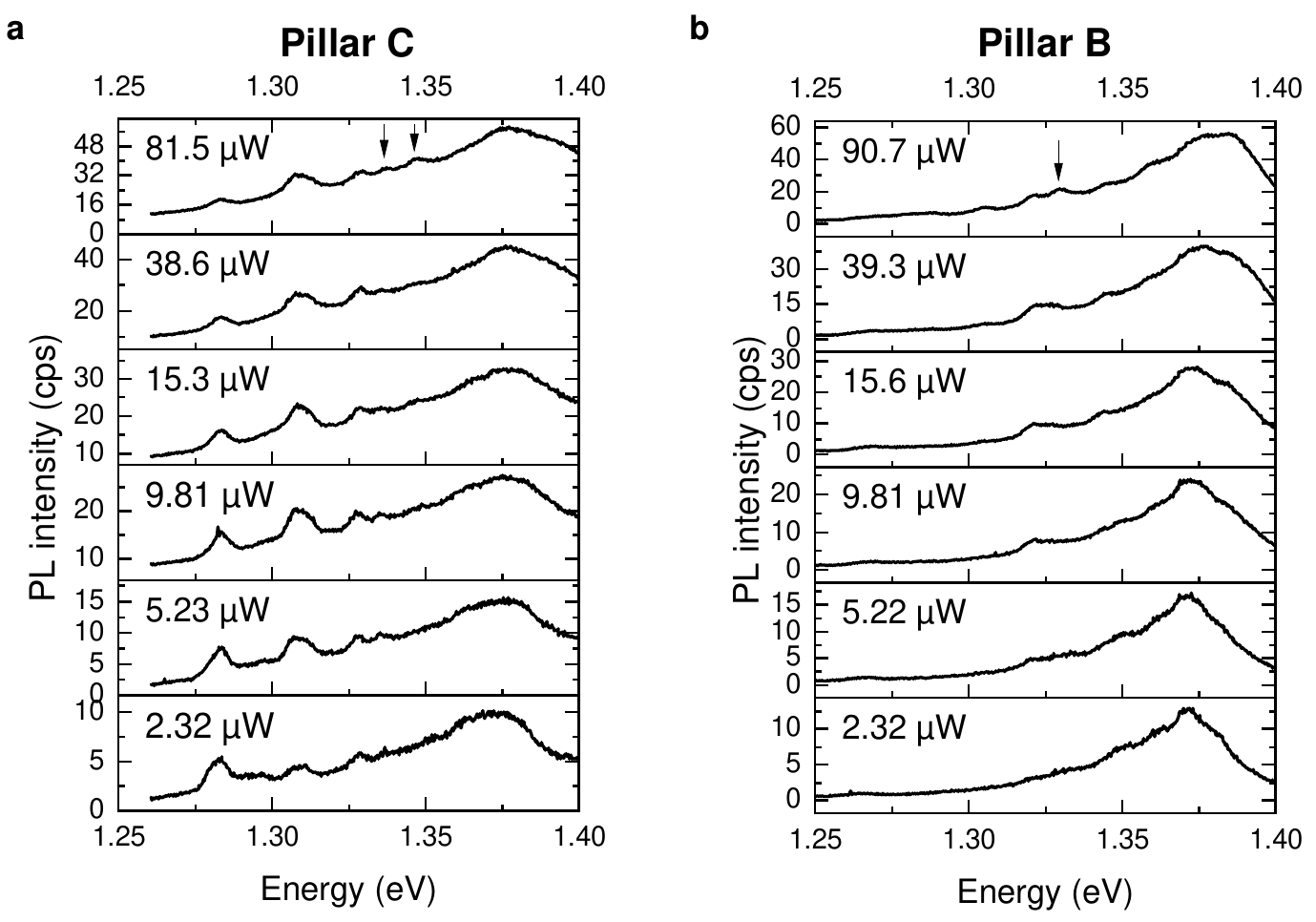}
\caption{\textbf{PL spectra at pillar sites C and B for different excitation powers.} ($T=\SI{10}{\kelvin}$, HeNe laser, focused to $\sim \SI{1}{\micro\meter}$)}
\label{fig:OtherPillars}
\end{figure*}

During our studies, pillar~A consistently showed the clearest signal with the smallest emission linewidths and we therefore focused on it for the quantitative analysis within the main manuscript. We present emission spectra of pillar~B~and~C for different excitation powers in Suppl.~Fig.~\ref{fig:OtherPillars}. They confirm the trend of the LIX emission being distributed towards higher-energy peaks for increasing excitation powers, which we saw in pillar~A. For pillar~C, one can see that the main LIX peak at low powers is at \SI{1.28}{\electronvolt}, whereas the peak at \SI{1.31}{\electronvolt} becomes more relevant at higher powers. Simultaneously, the two highest-power LIX peaks at \SI{1.337}{\electronvolt} and \SI{1.346}{\electronvolt} (indicated by arrows) only emerge as the excitation power increases. The spectra of pillar~B show the least clear trend. Nonetheless, a peak at \SI{1.33}{\electronvolt} (indicated by the arrow), blueshifted by \SI{9}{\milli\electronvolt} from the previous main LIX peak, appears towards the highest excitation powers.

\newpage


\subsection*{Supplementary Note 7: Polarization-resolved photoluminescence measurements}
\label{sub:LinPol}

\begin{figure*}[h!]
\includegraphics{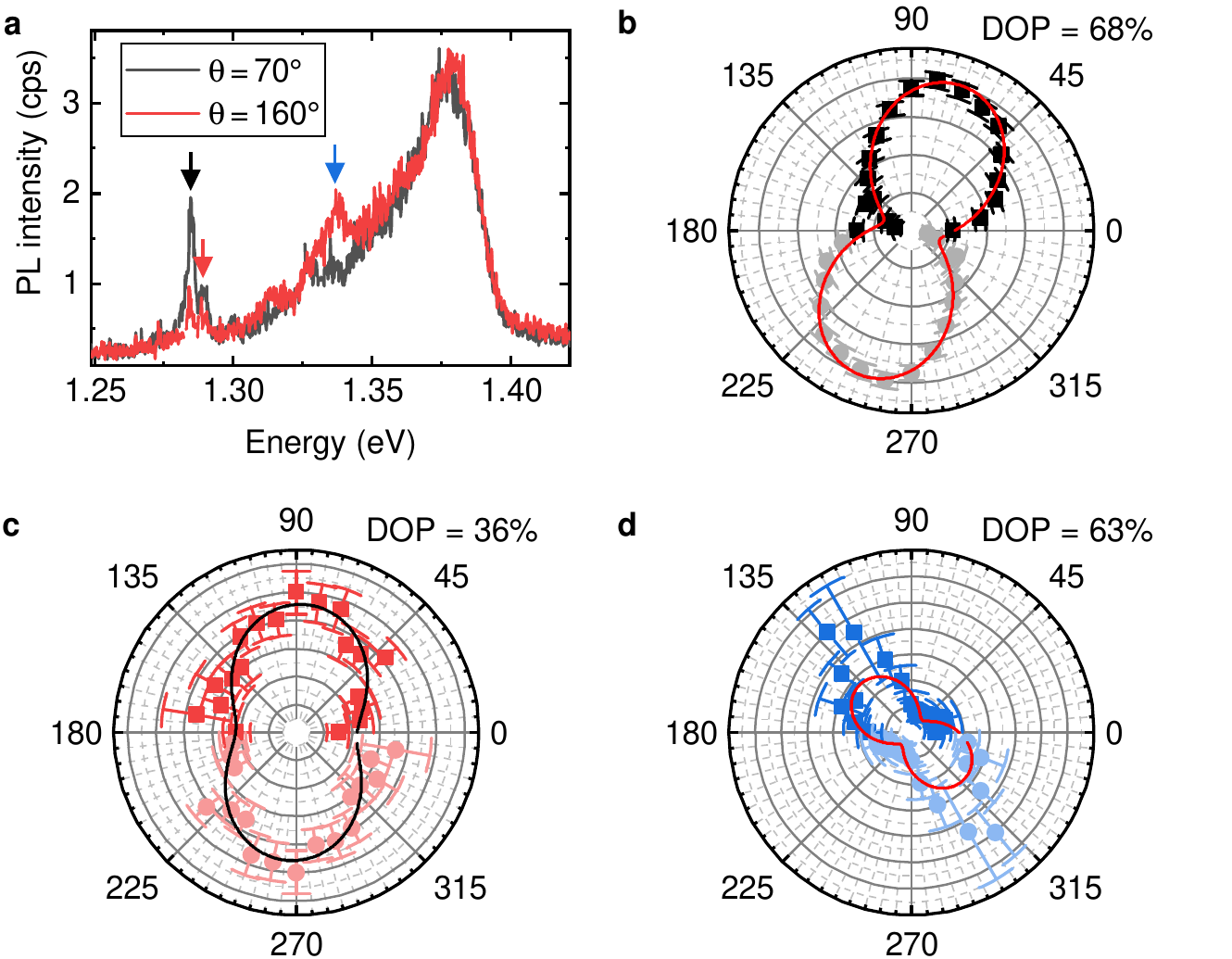}
\caption{\textbf{Linear-polarization-dependence of the LIX emission.} \textbf{a} PL spectrum at a pillar site for two examplary perpendicular polarizations. \textbf{b-d} Dependence of the LIX peaks marked in panel~a on the angle of the detected polarization. The faint points are mirrored from the actual dataset as a guide to the eye. The solid lines are fits to the data from which the degree of polarization is extracted.}
\label{fig:LinPol}
\end{figure*}

We analyzed the linear polarization of the emission of several LIX peaks. Two example spectra of the same pillar for perpendicular polarizations are shown in Suppl.~Fig.~\ref{fig:LinPol}a. While the free IX background does not change notably, the LIX peaks show strong intensity variations. The intensities of the marked peaks as a function of detected polarization angle are shown in Suppl.~Fig.~\ref{fig:LinPol}b-d. All peaks show preferential emission towards some angle. These angles do not coincide between the different peaks. The polarization of the excitation laser does not have any influence, ruling out valley coherence as the underlying effect. The highest degree of polarization (68\%) was observed for the lowest-energy peak. The reason for this polarization is not clear. We note that due to the sample geometry (heterobilayer forming a tent over a pillar) we may not probe the HBL in normal incidence.

\newpage


\subsection*{Supplementary Note 8: Enhanced excitation-power-dependent data for Fig.~2}
\label{sub:FullPower}

\begin{figure*}[h!]
\includegraphics{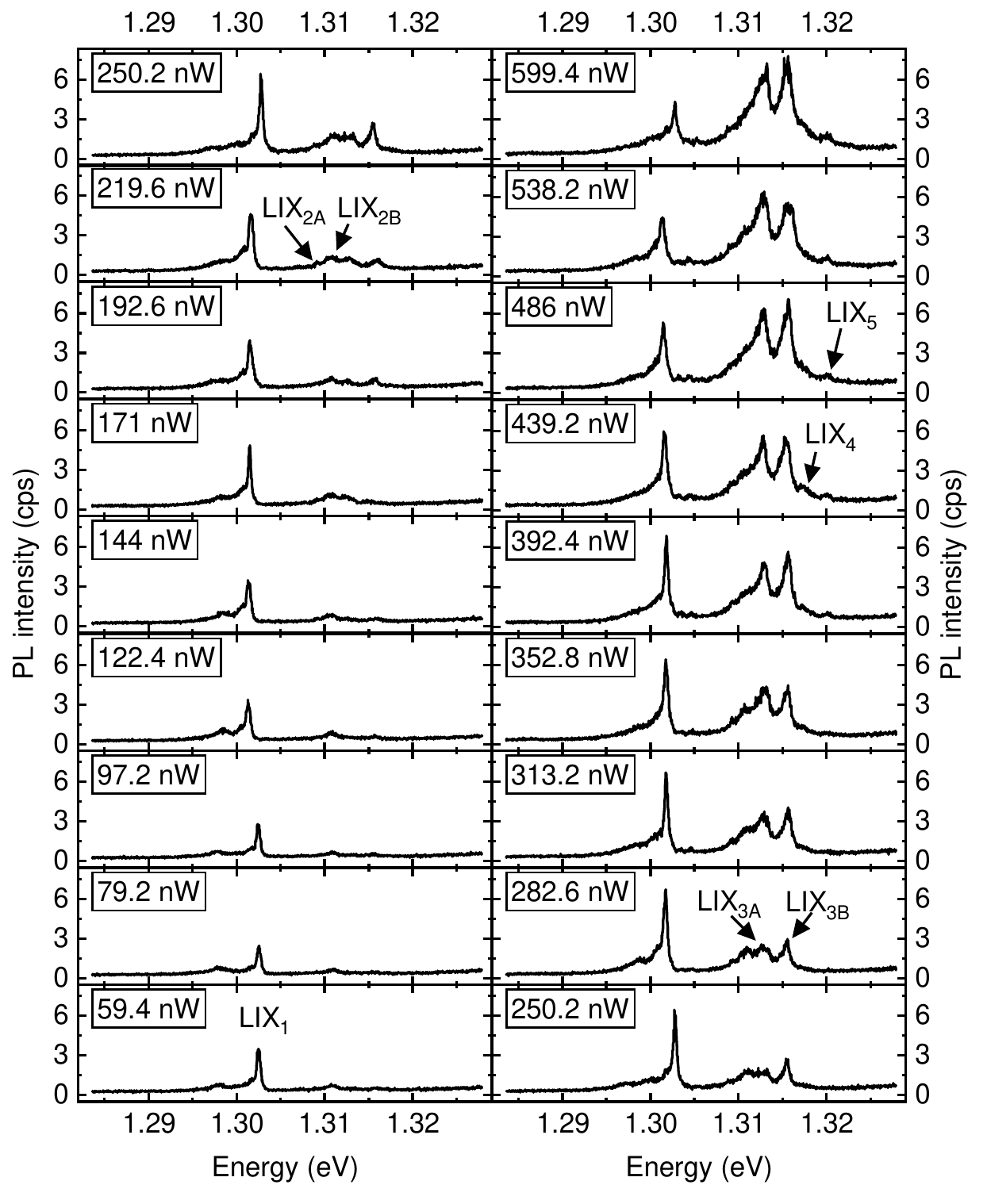}
\caption{\textbf{Additional exemplary LIX spectra for different excitation powers, from the dataset presented in Fig.~2a.}}
\label{fig:FullPower}
\end{figure*}

Supplementary~Figure~\ref{fig:FullPower} presents additional LIX PL spectra at varying excitation powers from the dataset presented in Fig.~2a. These data plus two additional spectra acquired at each of the presented powers were used to extract PL peak intensities and positions presented in Fig.~2b~and~c.

\newpage


\subsection*{Supplementary Note 9: Detailed power dependencies of multi-IX complexes}
\label{sub:PowerDetails}

\begin{figure}[h!]
\includegraphics{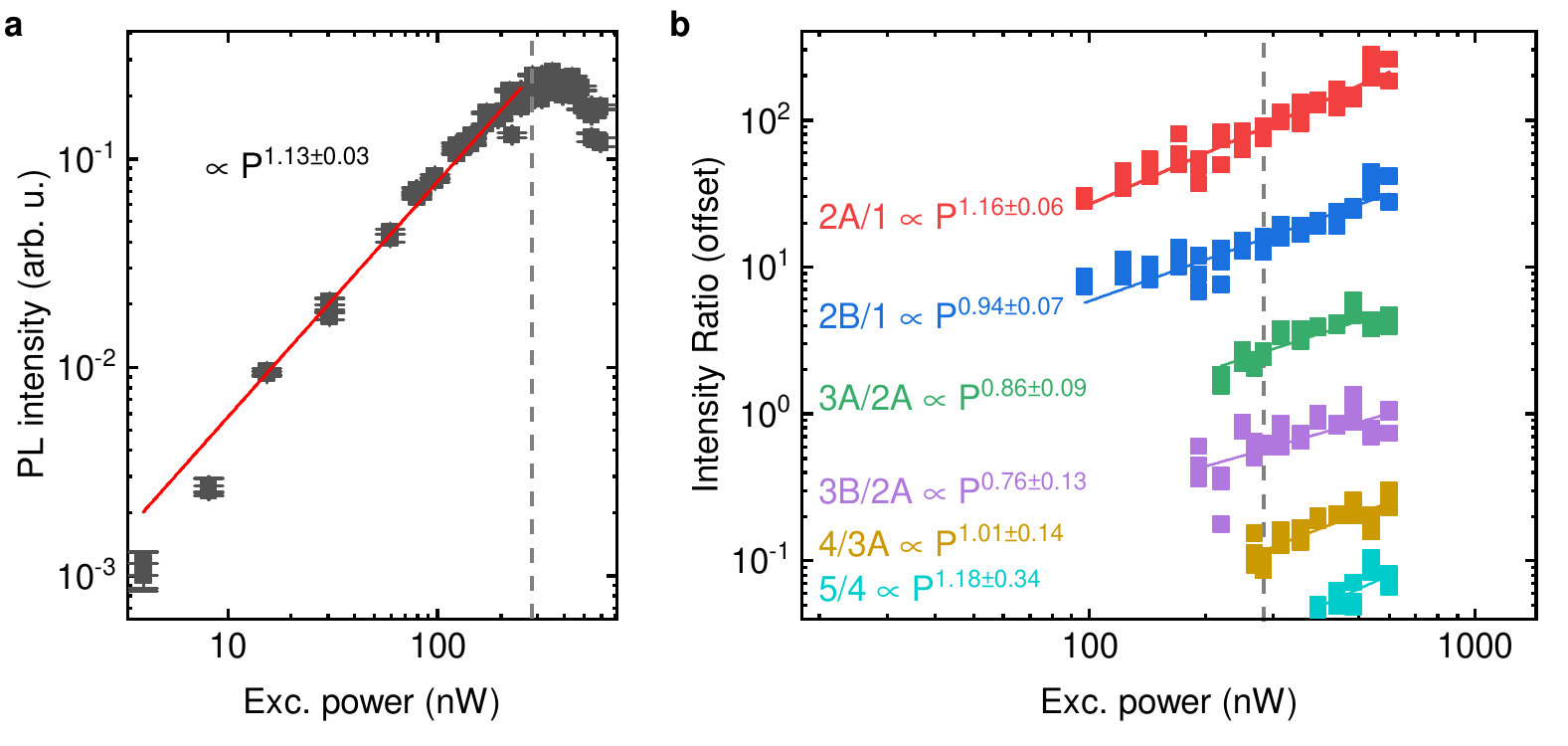}
\caption{\textbf{Excitation-power dependence of multi-IX peak intensities.} \textbf{a}, PL intensity of the LIX$_1$ peak, as presented in Fig.~2b, with additional data points for lower excitation powers. \textbf{b}, Selected PL intensity ratios of the few-IX peaks. The gray dashed lines in both panels mark the onset of saturation effects (same power as marked in Fig.~2b).}
\label{fig:PowerDetails}
\end{figure}

In order to verify our assignment of the single exciton peak (LIX$_1$ in Fig.~2), we performed PL measurements at even lower excitation powers. The integrated peak intensities in these measurements are presented in Suppl.~Fig.~\ref{fig:PowerDetails}a. Measurements at lower powers than the ones presented did not yield any usable signal. It can be seen that for almost two orders of magnitude in excitation power, the intensity increases linearly with the excitation before finally saturation effects occur. This is a clear indication that the LIX$_1$ stems from the recombination of a single exciton.

For some of the other peaks, we could only extract an intensity for few (or in the case of LIX$_5$ no) excitation powers before the onset of saturation. We used intensity ratios to substantiate our peak assignments with data from the saturated power regime. In order to justify this procedure, we consider a case in which excitons are randomly trapped in the potential at a rate $G$, which (like the exciton generation rate) is proportional to the excitation power, and decay with a radiative lifetime $\tau$. Both parameters could in principle depend on the occupation number $N$ of the trap. The occupation probabilities $n_N$ of the N-exciton states are then described by the following set of rate equations:

\begin{equation}
    \frac{d n_N}{dt} = \frac{(N+1)n_{N+1}}{\tau_{N+1}} - \frac{Nn_N}{\tau_{N}} + G_{N-1}n_{N-1} - G_{N}n_N
\end{equation}

If trapping and recombination rate are independent of N, the steady-state solution for these equations is known to be the Poissonian distribution:

\begin{equation}
    n_N = \frac{\left( G \tau \right)^N}{N!} e^{-G \tau}
\end{equation}

For a more general case, the exact distribution can take a different form. Focusing only on the ratio of populations of two states with occupation numbers $N+1$ and $N$, a quick, induction-based proof can show that:

\begin{equation}
    \frac{n_{N+1}}{n_{N}} = \frac{G_N \tau_{n+1}}{N+1} \propto P_{exc}
\end{equation}

This linear power dependence is robust against occupation-dependent capture and recombination rates and persists into the saturated regime (in contrast to the power law fits presented in Fig.~2b). We thus plot, as a function of excitation power, the ratios of the intensities of the multi-IX states with one state whose particle number we assume to be one below the examined state. This is shown in Suppl.~Fig.~\ref{fig:PowerDetails}b. In all cases, we obtain an approximately linear excitation-power dependence which is not influenced by the onset of saturation effects (marked by the gray dashed line), confirming our initial exciton number assignments.

\newpage


\subsection*{Supplementary Note 10: Calculations of the exchange energy of the localized interlayer biexciton}
We compare our experimental results for the energetic splitting of the observed biexciton
states to a theoretical model of exchange interactions between bilayer excitons as outlined in detail in Suppl.~Ref.~\cite{Bondarev2018x}. In this model, the tunnel exchange energy for interlayer biexciton complexes is computed in the absence of any external potential. The derivation is based on an Ansatz for the ground-state excitonic wavefunction of the (in-plane) relative electron-hole motion\cite{Leavitt1990x}. In our calculations, we adapted this model to take into account the presence of a weak symmetric harmonic confinement potential in order to model the strain-related trapping potential. For this, we added a harmonic trapping potential to the non-interacting part of the model Hamiltonian in Eq.~(2).

Two limiting cases can be considered for the trial wavefunction:
(i) For small interlayer distances $d$, the Hamiltonian resembles that of a two-dimensional hydrogen atom. The additional harmonic term then acts as an effective magnetic field that due to its low strength leaves the wavefunction unperturbed.
(ii) For large interlayer distances, the Coulomb interaction of electron and hole can be expanded in a Taylor series in terms of the in-plane distance $\rho$ of the two carriers. Upon truncating this expansion at second order, this gives rise to a term proportional to the square of the in-plane distance $\propto \rho^2$, which can be absorbed into the harmonic confinement potential introduced to the problem before. In doing so, we can closely follow the remaining steps of the derivation detailed in Suppl.~Ref.~\cite{Bondarev2018x} to evaluate the exchange energy of the interlayer biexciton.

Note that we implicitly assumed the validity of a couple of approximations for the derivation of binding and exchange energies presented in this work. Most notably, we assumed that (i) the effective electron and hole masses are equal, (ii) in-plane electrostatic interactions are well-described by Coulomb interaction potentials and (iii) image-charge effects are negligibly small. Further refining our model in order to take into account effects beyond the approximations (i)-(iii) is expected to result in more accurate results, of course. Despite these approximations, however, we expect the derived values in the main text to provide us with the correct orders of magnitude for the energy scales of interest. This is supported by the good agreement between theoretical predictions and performed measurements. Further note that, by construction, the employed variational Ansatz yields an upper limit for the exchange energy.

\newpage


\subsection*{Supplementary Note 11: Calculations of the binding energy of the localized multi-IXs}
\label{sub:bindingenergies}

\begin{figure}[h!]
\includegraphics{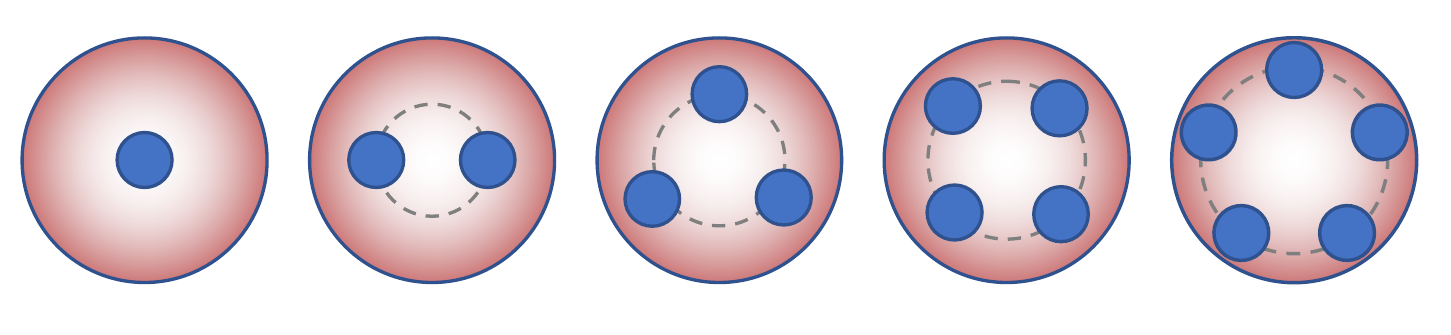}
\caption{\textbf{Illustrations of a few IXs (blue circles) compromising between drag towards the center of the trapping potential and mutual repulsion by sitting maximally spaced on an equipotential line of the trap.}}
\label{fig:Configurations}
\end{figure}

The binding energies of trapped multi-excitonic states were calculated classically in terms of point-like dipoles in an isotropic 2D harmonic potential. Note that we define positive binding energies as an increase of the multi-particle-state energy compared to the energy of the well-separated constituent particles. In order to minimize dipole-dipole repulsion, the IX were maximally spaced around an equipotential of the trapping potential. The constellations that arise from this condition are a line (biexciton), an equilateral triangle (triexciton), a square (quadexciton) and a regular pentagon (quintexciton), all centered around the minimum of the harmonic potential, as illustrated in Suppl.~Fig.~\ref{fig:Configurations}. As a function of the nearest-neighbor distance $\Delta\rho$, the harmonic and dipole-dipole potential energies are then given by:
\begin{eqnarray}
U^{2X} &=& 2 \cdot \frac{1}{2} \frac{\hbar^2}{M \ell^4} \left( \frac{\Delta \rho}{2} \right)^2 + 2 \frac{d^2}{4 \pi \epsilon} \left(\Delta \rho \right) ^{-3}\\
U^{3X} &=& 3 \cdot \frac{1}{2} \frac{\hbar^2}{M \ell^4} \left( \frac{\Delta \rho}{\sqrt{3}} \right)^2 + 6 \frac{d^2}{4 \pi \epsilon} \left(\Delta \rho \right) ^{-3}\\
U^{4X} &=& 4 \cdot \frac{1}{2} \frac{\hbar^2}{M \ell^4} \left( \frac{\Delta \rho}{\sqrt{2}} \right)^2 + 8 \frac{d^2}{4 \pi \epsilon} \left(\Delta \rho \right) ^{-3} + 4 \frac{d^2}{4 \pi \epsilon} \left(\sqrt{2} \Delta \rho \right) ^{-3}\\
U^{5X} &=& 5 \cdot \frac{1}{2} \frac{\hbar^2}{M \ell^4} \left( \frac{\sqrt{50+10\sqrt{5}}\Delta \rho}{10} \right)^2 + 10 \frac{d^2}{4 \pi \epsilon} \left(\Delta \rho \right) ^{-3} + 10 \frac{d^2}{4 \pi \epsilon} \left(\frac{1}{2} \left( 1 + \sqrt{5} \right) \Delta \rho \right) ^{-3}
\end{eqnarray}
, where $M$ is the total mass of the exciton and $\ell$ is the same effective confinement lengthscale defined in the main text. We minimize each of these expressions to find the equilbrium interparticle separation and equilibrium potential energy. Introducing $\tilde{E} = \left( \frac{d^2}{4 \pi \epsilon} \right)^{2/5} \left( \frac{\hbar ^2}{M \ell^4} \right)^{3/5}$, we find:
\begin{eqnarray}
U^{2X}_0 &=& \frac{5}{12^{3/5}} \tilde{E} \approx 1.125 \tilde{E}\\
U^{3X}_0 &=& \frac{5\cdot 18^{2/5}}{6} \tilde{E} \approx 2.648 \tilde{E}\\
U^{4X}_0 &=& \frac{5\left(8 + \sqrt{2} \right)^{2/5}}{2^{2/5}3^{3/5}} \tilde{E} \approx 4.806 \tilde{E}\\
U^{5X}_0 & \approx & 7.648 \tilde{E}
\end{eqnarray}
Combined with the energy of a single exciton inside the trap $E^X$, the transition energies of the multi-exciton states can then be expressed as:
\begin{eqnarray}
E^{2X \rightarrow 1X} &\approx& E^X + 1.125 \tilde{E}\\
E^{3X \rightarrow 2X} &\approx& E^X + 1.523 \tilde{E}\\
E^{4X \rightarrow 3X} &\approx& E^X + 2.158 \tilde{E}\\
E^{5X \rightarrow 4X} &\approx& E^X + 2.842 \tilde{E}
\end{eqnarray}
Due to the universal scaling with both, the strength of the harmonic potential and the dipole-dipole interaction, the ratios of blueshifts with respect to the single-exciton transition take fixed values:
\begin{eqnarray}
\frac{E^{3X \rightarrow 2X}-E^X}{E^{2X \rightarrow 1X}-E^X} &\approx& 1.354\\
\frac{E^{4X \rightarrow 3X}-E^X}{E^{2X \rightarrow 1X}-E^X} &\approx& 1.918\\
\frac{E^{5X \rightarrow 4X}-E^X}{E^{2X \rightarrow 1X}-E^X} &\approx& 2.526
\end{eqnarray}
As discussed in the main text, we observe the localized bi-, tri-, quad- and quintexciton emission to be blueshifted by \SI{8.4 \pm 0.6}{\milli\electronvolt}, \SI{12.4 \pm 0.4}{\milli\electronvolt}, \SI{15.5 \pm 0.6}{\milli\electronvolt} and \SI{18.2 \pm 0.8}{\milli\electronvolt} from the single exciton, respectively. Evaluating the blueshift ratios introduced above for these values, we obtain:
\begin{eqnarray}
\left( \frac{E^{3X \rightarrow 2X}-E^X}{E^{2X \rightarrow 1X}-E^X} \right)_{exp} &=& 1.48 \pm 0.12\\
\left( \frac{E^{4X \rightarrow 3X}-E^X}{E^{2X \rightarrow 1X}-E^X} \right)_{exp} &=& 1.85 \pm 0.15\\
\left( \frac{E^{5X \rightarrow 4X}-E^X}{E^{2X \rightarrow 1X}-E^X} \right)_{exp} &=& 2.17 \pm 0.18
\end{eqnarray}
The predictions from the model show good quantitative agreement, especially considering that the calculations are strictly classical and the exact trapping potential shape of our sample is not known. We emphasize that other spatial configurations of the excitons can easily be shown to be less favorable than the ones discussed above. For example, arranging three excitons in a line would result in $U^{3X}_{0,\textrm{line}} \approx 3.496 \tilde{E} > U^{3X}_0 $, a quadexciton with three excitons on a triangle and the fourth one in the center would give $U^{4X}_{0,\textrm{triangle}} \approx 5.493 \tilde{E} > U^{4X}_0 $ and a quintexciton consisting of four excitons on a square surrounding the fifth in the trap center would have $U^{5X}_{0,\textrm{square}} \approx 7.845 \tilde{E} > U^{5X}_0 $.

In addition to the energies of multi-excitonic states, the model also predicts equilibrium distances of the constituent excitons. These distances are not directly accessible in our experiments but we use the biexciton equilibrium distance $\Delta \rho _0 ^{2X} = \left( \frac{12 d^2 M \ell^4}{4 \pi \epsilon \hbar} \right)^{1/5}$ as $\Delta \rho$ in the calculations of the biexciton exchange energy.

\newpage


\subsection*{Supplementary Note 12: Thermal cycling}
\label{sub:ThermalCycling}

\begin{figure}[h!]
\includegraphics{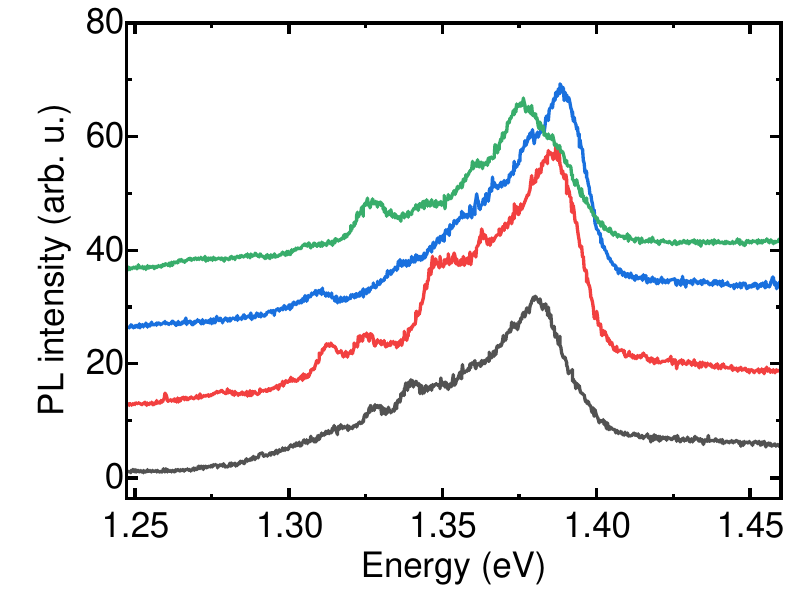}
\caption{\textbf{PL spectra obtained from the same pillar under the same excitation conditions ($T=\SI{10}{\kelvin}$, HeNe laser, \SI{30}{\micro\watt} focused to $\sim \SI{1}{\micro\meter}$) after thermal cycling.}}
\label{fig:ThermalCycling}
\end{figure}

We investigated the effect of thermal cycling on the localized IX by repeatedly cycling the sample temperature between \SI{10}{\kelvin} and \SI{300}{\kelvin}. Each time, we performed PL measurements at \SI{10}{\kelvin} under the same excitation conditions (cw excitation with a HeNe laser at \SI{633}{\nano\meter}, \SI{30}{\micro\watt} focused to a spot of \SI{1}{\micro\meter}). Supplementary~Figure~\ref{fig:ThermalCycling} compares typical emission spectra from the same pillar following successive cooldowns. We observe a clear change in the energy at which the LIX features appear. This effect was found to be even more pronounced than with typical strained WSe$_2$-monolayer samples tested under similar conditions. In order to mitigate the effects of thermal cycling, we performed the comprehensive power-dependent studies presented in the main manuscript (data in Fig.~2) with the sample placed inside a closed-cycle cryostat that facilitated all measurements during a single cool down. Under these conditions, we found the absolute energy of the LIX states to be stable over several weeks.

\providecommand{\latin}[1]{#1}
\makeatletter
\providecommand{\doi}
  {\begingroup\let\do\@makeother\dospecials
  \catcode`\{=1 \catcode`\}=2 \doi@aux}
\providecommand{\doi@aux}[1]{\endgroup\texttt{#1}}
\makeatother
\providecommand*\mcitethebibliography{\thebibliography}
\csname @ifundefined\endcsname{endmcitethebibliography}
  {\let\endmcitethebibliography\endthebibliography}{}

\end{document}